\documentclass[3p,preprint,10pt]{elsarticle}
\usepackage{geometry}
\setcitestyle{short&compress}
\newcommand*{\doi}[1]{\href{http://dx.doi.org/#1}{doi: #1}}
\usepackage{amsfonts}
\usepackage{amssymb}
\usepackage{pifont}
\usepackage[english]{babel}
\usepackage{graphicx}
\usepackage{xcolor}
\definecolor{darkblue}{rgb}{0.0,0.0,0.75}
\usepackage{bigints}
\usepackage{amsmath}
\usepackage{xfrac}
\usepackage{bm}

\usepackage{bm}

\usepackage{hyperref}
\hypersetup{
    pdftitle={Neural integration for constitutive equations using small data},    % title
    pdfauthor={Filippo Masi, Itai Einav},
    colorlinks=true,      
    linkcolor=blue,         
    citecolor=blue,     
    filecolor=blue,     
    urlcolor=blue
}
\usepackage[font=small,labelfont=bf]{caption}
\usepackage{subcaption}
\usepackage{caption}
\usepackage[modulo, displaymath, mathlines]{lineno}
\let\oldequation\equation
\let\oldendequation\endequation

\renewenvironment{equation}
  {\linenomathNonumbers\oldequation}
  {\oldendequation\endlinenomath}
  
\makeatletter
\def\ps@pprintTitle{%
\let\@oddhead\@empty
 \let\@evenhead\@empty
 \def\@oddfoot{}%
\let\@evenfoot\@oddfoot
 }
\makeatother

\usepackage{xfrac}
\usepackage[algoruled,shortend,vlined]{algorithm2e}
\SetAlCapNameFnt{\small}
\SetAlCapFnt{\small}

\usepackage{mathrsfs}

\usepackage{titlesec}

\usepackage{float}
\usepackage{booktabs}
\usepackage{makecell}  % new
\usepackage{tabularx}
\usepackage{siunitx}

\newcolumntype{C}[1]{>{\centering\arraybackslash}m{#1}}
\usepackage[most]{tcolorbox}

\tcbset{colback=yellow!20!white, colframe=black, boxrule=1pt,
        highlight math style= {enhanced, %<-- needed for the ’remember’ options
            colframe=red,colback=red!10!white,boxsep=0pt}
        }
\usepackage{accents}
\usepackage{scalerel,stackengine}
\def\sq{\mathbin{\scalerel*{\strut\rule[-.2ex]{1.5ex}{1.5ex}}{\cdot}}}

\usepackage{etoolbox}
\patchcmd{\MaketitleBox}{\footnotesize\itshape\elsaddress\par\vskip36pt}{\footnotesize\itshape\elsaddress\par\parbox[b][36pt]{\linewidth}{\vfill\hfill\textnormal{\today}\hfill\null\vfill}}{}{}%
\patchcmd{\pprintMaketitle}{\footnotesize\itshape\elsaddress\par\vskip36pt}{\footnotesize\itshape\elsaddress\par\parbox[b][36pt]{\linewidth}{\vfill\hfill\textnormal{\today}\hfill\null\vfill}}{}{}%

\let\today\relax

\AtBeginDocument{\setlength\abovedisplayskip{1pt}}
\AtBeginDocument{\setlength\belowdisplayskip{1pt}}
\titlespacing\section{0pt}{12pt plus 4pt minus 2pt}{4pt plus 2pt minus 2pt}
\titlespacing\subsection{0pt}{12pt plus 4pt minus 2pt}{4pt plus 2pt minus 2pt}
\titlespacing\subsubsection{0pt}{12pt plus 4pt minus 2pt}{2pt plus 2pt minus 2pt}

\titleformat*{\subsection}{\bfseries}
\titleformat*{\subsubsection}{\normalshape}

\begin{document}

\begin{frontmatter}
\title{\Large{\bfseries Neural integration for constitutive equations using small data}}
\author[1]{Filippo Masi\corref{cor1}}
\ead{filippo.masi@sydney.edu.au}
\author[1]{Itai Einav}
\ead{itai.einav@sydney.edu.au}
\cortext[cor1]{Corresponding author}
\affiliation[1]{organization={
Sydney Centre in Geomechanics and Mining Materials,
  School of Civil Engineering,
  The University of Sydney},
postcode={NSW 2006},
city={Sydney},
country={Australia.}}
\begin{abstract}
Data-driven models based on deep learning algorithms intend to overcome the limitations of traditional constitutive modelling by directly learning from data. However, the need for extensive data that collate the full state of the material is hindered by traditional experimental observations, which typically provide only \emph{small data} -- sparse and partial material state observations.
To address this issue, we develop a novel deep learning algorithm referred to as \emph{Neural Integration for Constitutive Equations} to discover constitutive models at the material point level from scarce and incomplete observations. It builds upon the solution of the initial value problem describing the time evolution of the material state, unlike the majority of data-driven approaches for constitutive modelling that require large data of increments of state variables. Numerical benchmarks demonstrate that the method can learn accurate, consistent, and robust constitutive models from incomplete, sparse, and noisy data collecting simple conventional experimental protocols.
\vspace{6pt}

\noindent \textbf{Keywords} \noindent Deep learning; Constitutive modelling; Small data; Thermodynamics.
\end{abstract}
\end{frontmatter}

\section{INTRODUCTION}
\noindent Increase in computational power and large availability of data has recently brought to a transformative focus in mechanics towards data-driven models. Mainly based on deep learning algorithms, these approaches represent a promising set of tools for addressing open challenges in solid mechanics and material science (see, without being exhaustive, \citep{ghaboussi1998new,lefik2003artificial,yun2008new,masi2021thermodynamics,yin2022interfacing,flaschel2023automated,cueto2023thermodynamics,peirlinck2024automated,linden2023neural,tran2020active}).
Within this framework, data-driven approaches have increasingly been explored to overcome the limitations associated with the traditional constitutive modelling paradigm that builds upon the development of material models, which tend to follow iterative calibrations and ongoing adjustments of evolution equations of state variables \citep{coleman1967thermodynamics,maugin2015saga}. This process usually relies on only a few experimental observations for any new given material. Instead, data-driven approaches aim to learn directly from observations, thus overcoming the necessity for subjective human's trial-and-error calibration (for a review, we refer to, \citep{peng2021multiscale,dornheim2023neural}).

Further motivated by the need of developing models that respect established physical principles, recent developments have given rise to a wealth of methods we now refer to as physics-informed machine learning (see \citep{karniadakis2021physics}). This new class of data-driven approaches enables to leverage prior knowledge stemming from physics, such as thermodynamics, by devising appropriate architectures that enforce constraints, resulting in enhanced generalisation and extrapolation capabilities (see \emph{e.g.} \citep{masialert}).

Nonetheless, at present, only a few approaches can directly learn material models from the actual set of data  experimental observations tend to provide (for instance \citep{tasdemir2022strategy,flaschel2023automated,wang2023automated,xu2023small}). The main reason stems from the fact that data-driven methods generally demand \emph{big data} for learning. Here, the \emph{big} refers to the set of information that collects extensive observations of the full state of the system (density, stress, temperature, etc.) as well as the numerical details of the loading protocol. These are necessary ingredients for the vast majority of current `incremental formulations' that tend to require full experimental data about the increments of all the state variables (for more detail, we refer to Section \ref{sec:theory}). Despite the ongoing development of experimental techniques that aim at better characterising materials (see, for instance, \citep{tengattini2023micromechanically}), those techniques can still not thoroughly and fully capture the complete state of materials at any loading scenario, especially complex ones. We are thus confronted with the big problem of \emph{small data}, where \emph{small} refers here to a few partial (incomplete) sets of measurements (see also \citep{xu2023small}.

To overcome the aforementioned difficulties, the conventional strategy (among others, \emph{cf.} \citep{xu2023small}) consists of performing \emph{in silico} experiments where all necessary information can be generated through computer simulations of idealised materials (see \emph{e.g.} \citep{logarzo2021smart,yin2022interfacing,vlassis2022geometric,masi2023evolution,ma2023learning}). However, the underlying drawback of this approach dwells from the need of knowing both the physics and the geometries at the fine material scale (microstructure), which is practically impossible for complex, opaque materials and boundaries, especially during dynamic loading. Recently, a new class of approaches arose to address the lack of knowledge of the full state of the materials (referred to as EUCLID, \citep{flaschel2021unsupervised,flaschel2023automated}). The method resorts to the automatic identification of constitutive models based on iterative solutions of the boundary value problem describing the deformation of the specimen, during an experiment. Whilst this approach enables to discover constitutive equations from direct measurement data, it still requires high-resolution information of internal full-field displacement data, which may not be readily available, especially for opaque materials with complex microstructures, such geo- and bio-materials.\\

\noindent The key point of the current contribution is the development of a novel deep learning method for the discovery of constitutive equations, at the material point level, from small data -- that is, from partial and sparse observations of the material state. A core idea is in resorting to established practices in constitutive modelling. Traditionally, the evaluation of a constitutive model against experimental evidence follows the solution of an initial value problem: after specification of the initial values for the state variables, one integrates in time the evolution equations, computes the corresponding stress, and compares with experimental observations, relying solely on available stress-strain data pairs. This work embraces part of this long-standing practice as it proposes to discover constitutive models based on the solution of the initial value problem associated with the evolution equations. Thus, there is no need of computing incremental values or rates of state variables, which are not directly available in experiments, or knowing the full material state at each time (as opposed to incremental formulations).

\begin{figure}[h]
\centering
\includegraphics[width=0.7\textwidth]{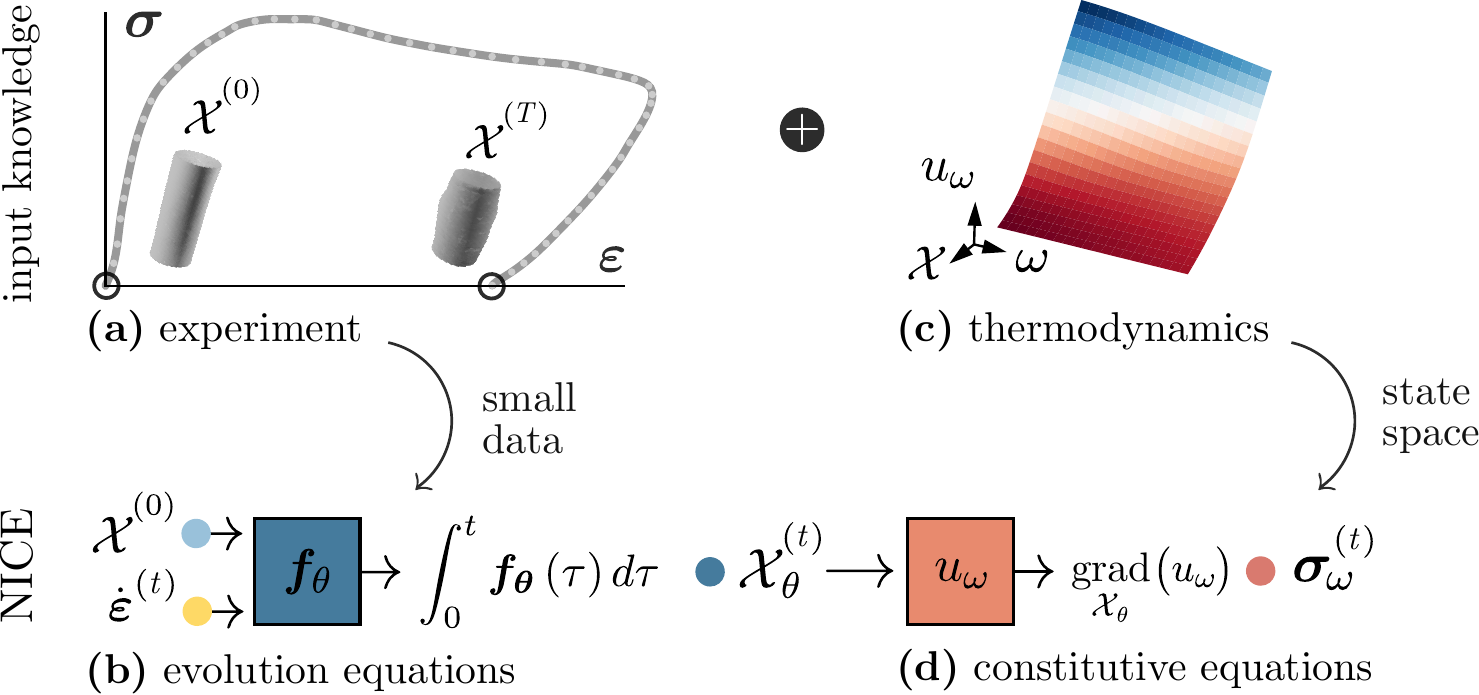}
\caption{Graphical illustration of NICE: \emph{Neural Integration for Constitutive Equations}. Experimental tests (a) gather the stress-strain ($\bm{\sigma}$ versus $\bm{\varepsilon}$) response, as well as initial and final material configurations, represented here by an incomplete set of sampled state variable data ($\mathcal{X}^{(0)}$ and $\mathcal{X}^{(T)}$). From these measurement data, the system learns the evolution equations (b) through a neural differential operator $\bm{f}_{\bm{\theta}}$ with parameters $\bm{\theta}$. The continuous-time and integral form of $\bm{f}_{\bm{\theta}}$ then computes the state space trajectory $\smash{\mathcal{X}_{\bm{\theta}}^{(t)}}$ for the stress evaluation (d), using a second network parametrised by $\bm{\omega}$ that represents the specific internal energy $u_{\bm{\omega}}$, along with constitutive constraints derived from thermodynamic principles (c).
}
\label{fig:intro}
\end{figure}

To achieve this, we leverage the framework provided by neural differentials equations \citep{chen2018neural}. Neural differential equations are a deep learning algorithm that blends artificial neural networks with time integration and enable learning time-continuous dynamics in the form of a system of differential equations solely from discrete-time observations. The framework constitutes the basis of our new approach, to which we refer in the following as the \emph{Neural Integration for Constitutive Equations} (NICE) method, see Figure \ref{fig:intro}. 

Neural integration of constitutive equations replaces the classical incremental data-fitting approach with a time integral formulation, relying on the computation of gradients with respect to network parameters using the adjoint sensitivity method \citep{pontryagin1987mathematical,chen2018neural}. The learning process thus resorts to the minimisation of the error of predicted trajectories with respect to sparsely sampled measurement data. In addition, our approach leverages previous developments of the Thermodynamics-based Artificial Neural Networks (TANN, see \citep{masi2021thermodynamics, masi2023evolution}), where the material is solely defined by a set of state variables (following the seminal work of \citep{coleman1967thermodynamics}). Here we extend the theoretical formalism to account for the dependence of the material behaviour on the density, which allows us to properly model porous materials.
The thermodynamics, on one hand, and the integral formulation, on the other hand, make the newly proposed method not only capable to identify consistent material representations, but also learning from very scarce and incomplete data.

The performance and accuracy of the new method are investigated using numerical benchmarks, as these allow a thorough probing of the predictions by direct comparison with a reference constitutive model. With the ultimate goal of applying the proposed approach in real scenarios, we focus our attention on the way the training data sets are generated. To this end, we simulate numerical experiments and collect data by respecting conventional protocols, marked by the application of monotonic loading and subsequent unloading until the specimen reaches an unstressed state, rather than including cyclic or more exotic loading paths, as these are rarely probed in real experiments.

The approach is tested by considering two emblematic benchmarks. The first one involves an elasto-plastic material and represents the proof of principle in learning from partial data -- that is, without using as an input elastic strain data. Thanks to the integral formulation and thermodynamics, the neural model identifies appropriate initial guesses for the values of the elastic strain which are then part of the whole learning process. Their evolution is predicted by only fitting the time evolution of the stress.

The second benchmark consists instead of a porous, elasto-plastic material, described by three state variables: density, elastic strain, and solid fraction. While the evolution of density is directly prescribed by the balance of mass, the evolution of the solid fraction is of completely different nature with respect to the elastic strain. This second example showcases the ability of the method to learn from sparse and limited data. In particular, we consider both a dense and a sparse time acquisition of the state space, and assess the predictions at inference for unobserved loading paths and protocols. An overall good agreement is found, even in the case when only two samples of the solid fraction are available: at the beginning and at the end of each loading path.

With the aim of evaluating the influence of corrupted measurement data, we further consider synthetic noise in the values of all quantities of interest (in contrast, for instance, with \citep{masi2021thermodynamics}). The approach demonstrates high level of robustness to noise in the training data.\\

\noindent The paper is structured as follows. Section \ref{sec:theory} outlines a few background elements upon which our approach is built. Section \ref{sec:NDCE} presents the methodology of the neural integration for constitutive equations. Here, we focus attention on both the practical implementation and the scalability of the methodology in dealing with scarce, limited data, sampled at irregular times. Section \ref{sec:benchmarks} showcases the accuracy, performance, and robustness of the proposed methodology by focusing on two rather simple applications, yet with powerful implications for future extensions to experimental observations. In parallel, we present in \hyperref[appendixA]{Appendix} a detailed comparative analysis between the time integral formulation and the conventional incremental (or rate) formulation. Although the latter is rooted in the same thermodynamic framework, it exhibits increased sensitivity to noise and is incapable of discovering constitutive models when faced with incomplete or sparse data sets. Finally, Section \ref{sec:conclusions} discusses future developments that might allow the field to address the inherent difficulties associated with direct learning from real experimental data.

The following notation is used throughout the manuscript: $\mathbf{a}\cdot \mathbf{b} = a_i b_i$, $\bm{\sigma}:\dot{\bm{\varepsilon}}=\sigma_{ij} \dot{\varepsilon}_{ij}$, $\boldsymbol{1}$ is the identity tensor, $\mathrm{tr}(\cdot)$ represents the trace of a second-order tensor, and $\mathrm{div}\,\mathbf{v} = \frac{\partial v_i}{\partial x_i}$, with $x_i$ the spatial coordinates, $v_j$ a vectorial quantity, and $i,j=1,2,3$. Einstein's summation is implied for repeated indices. For the ease of reading, we distinguish functions from their values with a superposed caret, whenever a marked difference is important: for instance, $\widehat{u}$ is a function and $u$ its value.

\section{THEORETICAL BACKGROUND}
\label{sec:theory}
\noindent We start by introducing the theoretical background behind the core of the new method: (1) the commonly used structure of data-driven neural networks in the field of constitutive modelling and the corresponding caveats, (2) the mathematical framework that enables overcoming them, and (3) the constitutive restrictions stemming from thermodynamics.

\subsection{The common structure: incremental formulation}
\label{par:incremental}
\noindent The vast majority of data-driven deep learning approaches for constitutive modelling builds upon an `incremental formulation' of the evolution equations. Many architectures and structures have been deployed with this purpose, and mainly feed-forward, residual, and recurrent neural networks (\emph{e.g.} \citep{ghaboussi1998new,lefik2003artificial,yun2008new,masi2021thermodynamics,
wu2022recurrent,vlassis2022geometric,wang2023automated,masi2023evolution}).

For compactness of notation, we avoid specifying the state space of the material at this point (further introduced in subsection \ref{par:thermodynamics}) and consider a hypothetical time-evolving system whose state is denoted by $\mathbf{y}$. The aforementioned approaches describe the evolution of the state according to the incremental formulation
\begin{equation}
{\mathbf{y}^{(n)} = \mathbf{y}^{(n-1)}} + \bm{h}_{\bm{\theta}} \left( \mathbf{y}^{(n-1)}, \mathbf{q} \right),
\label{eq:rateform}
\end{equation}
where $\mathbf{y}^{(n)}$ is the value of the state at discrete times $\{t^n\}_{n=1}^N$; $\bm{h}_{\bm{\theta}}$ is a generic neural operator with parameters $\bm{\theta}$ that establishes the evolution of the state; and $\mathbf{q}$ represents the set of additional control variables the state depends upon, such as time or the values of the state at preceding times. Depending on the specified control variables, expression (\ref{eq:rateform}) can represent different neural network architectures. Note that the extension of (\ref{eq:rateform}) to a incremental formulation is also possible relying on finite-difference approximations of the rates of $\mathbf{y}$ (see for instance \citep{masi2023evolution}). 

Due to the dependence on the time interval discretisation, training the operator $\bm{h}_{\bm{\theta}}$ requires dense, time-parametrised data that describe the full state. This is especially true when dealing with states that vary rapidly in time, due to the incapacity of modelling the instantaneous rate of change in $\mathbf{y}$. Additional limitations lie on the need of having regularly spaced intervals (see \citep{chen2018neural}) and on the impossibility of operating with incomplete measurement data. When only partial knowledge of the state $\mathbf{y}^{(n-1)}$ is available, it is impossible to infer the next state $\mathbf{y}^{(n)}$.

Reformulating in the context of evolution equations, the incremental form (\ref{eq:rateform}) cannot yield accurate representations of evolving state variables in the experimental setting of complex materials, see Figure \ref{fig:intro}.

\subsection{Neural differential equations: integral formulation}
\label{par:neural}
\noindent In contrast with the aforementioned class of incremental formulations, neural differential equations (see \citep{weinan2017proposal,haber2017stable,chen2018neural}, among others) are deep learning algorithms allowing the parametrisation of continuous-time dynamics of the state ${\mathbf{y}}(t)$ formulated as an ordinary differential equation in the form
\begin{equation}
\dot{\mathbf{y}} = \dfrac{d{\mathbf{y}}}{dt}(t) = \bm{f}_{\bm{\theta}}\left(\mathbf{y}(t),\mathbf{q} \right), \quad {\mathbf{y}}(0)=\mathbf{y}^{(0)}, \quad t \in \left[0,T\right]
\label{eq:neuralODE}
\end{equation}
where $\mathbf{q}$ are the control variables and $\bm{f}_{\bm{\theta}}$ is an evolution equation parametrised by a feed-forward neural network, where $\bm{\theta}$ collects the weights and the bias terms and under the premises that the operator $\bm{f}_{\bm{\theta}}$ is integrable.
 
From the knowledge of the initial condition $\mathbf{y}^{(0)}$, the state at any time $t$ is obtained by solving the initial value problem (\ref{eq:neuralODE}) with a differential equation solver
\begin{equation}
{\mathbf{y}}_{\bm{\theta}}(t) = \mathbf{y}^{(0)}+\int_{0}^t \bm{f}_{\bm{\theta}}\big(\mathbf{y}(\tau);\mathbf{q} \big) \; d \tau =  \underset{\left(0, t\right]}{\mathrm{ODESolve}}\left(\mathbf{y}^{(0)},\bm{f}_{\bm{\theta}}, \mathbf{q}\right).
\label{eq:neuralODE2}
\end{equation}
The parameters $\bm{\theta}$ are determined by means of the gradient descent algorithm \citep{geron2019hands}, using the adjoint sensitivity method \citep{pontryagin1987mathematical}. This passes through the minimization of the error, at discrete-time points $t^{n}$ with $n=1, \ldots, N$ and $t^N\equiv T$, between the neural network predictions, ${\mathbf{y}}_{\bm{\theta}}(t^n)$, and the set of measurement data, $\mathbf{y}^{(n)}$, using \emph{e.g.}, the mean squared error,
\begin{equation}
\mathcal{L} = \dfrac{1}{N} \sum_{n = 1}^{N}\bigg( \underset{ \left(0, t^{n}\right]}{\mathrm{ODESolve}}\left(\mathbf{y}^{(0)},\bm{f}_{\bm{\theta}}, \mathbf{q}\right) -\mathbf{y}^{(n)} \bigg)^2.
\end{equation}
The formulation (\ref{eq:neuralODE}-\ref{eq:neuralODE2}) can be further extended to model a wide class of differential equations, and not only ordinary differential equations (see \emph{e.g.} \citep{dupont2019augmented,massaroli2020dissecting,kidger2022neural}).

Note that the idea of employing neural differential equations for constitutive modelling has not been properly addressed so far (a few formulations are presented in \citep{jones2022neural,tac2022data,ma2023learning}, however these works do not address the problem of small, incomplete measurement data which is the motivation of the present work).

\subsection{Thermodynamics and constitutive restrictions}
\label{par:thermodynamics}
\noindent We consider a porous, single component material, undergoing infinitesimal strains and isothermal processes. The state of the material is assumed to be completely defined by a set of $p+2$ state variables $\mathcal{X} = \{ \rho, \bm{\varepsilon}^e, \mathbf{z}_1, \ldots \mathbf{z}_{p}\}$, where $\rho$ is the volumetric mass density, $\bm{\varepsilon}^e$ is the elastic strain, and $\mathbf{z}_i$, with $i=1, \ldots, p$, are (additional) dissipative state variables describing irreversible phenomena (see \citep{einav2023hydrodynamics}), such as damage and dislocation density. For compactness of notation, we denote the set of dissipative state variables as $\mathbf{z} =\{\mathbf{z}_1, \ldots \mathbf{z}_{p}\}$. 

Alternatively, some often consider the total strain $\bm{\varepsilon}$, rather than the elastic one. However, here we disregard this second path as a classification based on the total strain renders the state space itself dependent of the (arbitrarily chosen) referential configuration (see \citep{rubin2001physical}).

By virtue of the classification $\{ \rho, \bm{\varepsilon}^e, \mathbf{z}\}$ and the axiom of equipresence \citep{truesdell1960classical}, the specific (per unit volume) internal energy is specified as
\begin{equation}
u \equiv \widehat{u}\left(\rho,\bm{\varepsilon}^e,\mathbf{z}\right),
\label{eq:energy}
\end{equation}
where the superposed caret in $\widehat{u}$ serves to distinguish the internal energy function from its values.

Constitutive equations must comply with the energy and entropy balance (\emph{i.e.} first and second law of thermodynamics). These principles are here formulated in terms of the dissipation inequality (Clausius-Duhem inequality, \citep{coleman1967thermodynamics}), namely
\begin{equation}
d = \bm{\sigma}:\dot{\bm{\varepsilon}} - \left(\dot{u} - \dfrac{u}{\rho}\dot{\rho} \right) \ge 0,
\label{eq:clausius_}
\end{equation}
where $d$ is the specific mechanical dissipation rate, $\bm{\sigma}$ is the stress tensor, $\bm{\varepsilon}$ and $\dot{\bm{\varepsilon}}$ are the infinitesimal strain and strain rate tensors. A more general formulation could follow Landau's hydrodynamic principles, which could allow one to consider other scales of temperature and viscous stress (see \citep{landau2013statistical,jiang2009granular,einav2023hydrodynamics}), but this goes beyond the focus of the current paper.

Further assuming the internal energy to be differentiable with respect to the state variables, the time derivative in inequality (\ref{eq:clausius_}) reads
\begin{equation}
\dot{u} = \dfrac{\partial \widehat{u}}{\partial \rho} \dot{\rho}+\dfrac{\partial \widehat{u}}{\partial \bm{\varepsilon}^e} \dot{\bm{\varepsilon}}^e + \dfrac{\partial \widehat{u}}{\partial \mathbf{z}} \cdot \dot{\mathbf{z}}.
\label{eq:energy_rate}
\end{equation}

\noindent We further define the plastic strain rate $\dot{\bm{\varepsilon}}^p$, such that
\begin{equation}
\dot{\bm{\varepsilon}}^p \equiv \dot{\bm{\varepsilon}}-\dot{\bm{\varepsilon}}^e.
\label{eq:decomp}
\end{equation}
From the mass balance, \emph{i.e.}, $\dot{\rho}+\rho\,\mathrm{div}\,\mathbf{v} = 0$, with the spatial velocity given by $\mathbf{v} = \frac{d \mathbf{x}}{dt}$, we obtain the evolution equation of the volumetric mass density,
\begin{equation}
\dfrac{\dot{\rho}}{\rho} = \boldsymbol{1}: \dot{\bm{\varepsilon}},
\label{eq:mass}
\end{equation}
where we have adopted the usual sign convention in soil mechanics that consider positive deformation in compression, and used $\boldsymbol{1}$ as the identity tensor.

Finally, substitution of the time derivative (\ref{eq:energy_rate}) and expressions (\ref{eq:decomp}-\ref{eq:mass}) in the Clausius-Duhem inequality (\ref{eq:clausius_}) yields
\begin{equation}
\Bigg ( \bm{\sigma} - \dfrac{\partial \widehat{u}}{\partial \bm{\varepsilon}^e} -\left( \rho \dfrac{\partial \widehat{u}}{\partial \rho} -u\right) \boldsymbol{1} \Bigg ):\dot{\bm{\varepsilon}}^e+\Bigg ( \bm{\sigma} +\left( \rho \dfrac{\partial \widehat{u}}{\partial \rho} -u \right) \boldsymbol{1} \Bigg ):\dot{\bm{\varepsilon}}^p-\dfrac{\partial 
\widehat{u}}{\partial \mathbf{z}}\cdot \dot{\mathbf{z}}-d = 0.
\end{equation}
Considering the arbitrariness of $\dot{\bm{\varepsilon}}^e$ and $\dot{\mathbf{z}}$ and that the dissipation rate does not depend on $\dot{\bm{\varepsilon}}^e$ brings to the following constitutive restrictions
\begin{subequations}
\label{eqs:stress_dissipation}
\begin{align}
\label{eq:stress}
\bm{\sigma}& = \bm{\sigma}^e+p^T \boldsymbol{1},\\
\label{eq:dissipation}
d &= \bm{\sigma}^e:\dot{\bm{\varepsilon}}^p-\bm{\tau}\cdot \dot{\mathbf{z}} \geq 0,\\
\label{eq:definition}
\bm{\sigma}^e &\equiv \dfrac{\partial \widehat{u}}{\partial \bm{\varepsilon}^e}, \quad p^T \equiv \rho \mu-u, \quad \mu \equiv \dfrac{\partial \widehat{u}}{\partial \rho}, \quad \bm{\tau} \equiv \dfrac{\partial \widehat{u}}{\partial \mathbf{z}}
\end{align}
\end{subequations}
where $\bm{\sigma}^e$ is the elastic stress, conjugate to $\bm{\varepsilon}^e$; $\mu$ is the chemical potential, conjugate to $\rho$; $\bm{\tau}$ are the thermodynamic forces of the dissipative mechanisms, conjugate to $\mathbf{z}$; and $p^T$ is the thermodynamic pressure (see \citep{einav2018hydrodynamic}).

Constitutive equations (\ref{eqs:stress_dissipation}) are accompanied by the evolution equations of the state variables that, in the most general form, read 
\begin{equation}
\label{eq:evol}
\dot{\mathcal{X}}=\bm{f}\left(\mathcal{X}, \dot{\bm{\varepsilon}}\right), \qquad 
{\bm{f}} \equiv \big\{\bm{f}^{\rho},
\bm{f}^{\bm{\varepsilon}^e},
\bm{f}^{\mathbf{z}}\big\},
\end{equation}
which has to fulfil the dissipation inequality (\ref{eq:clausius_}), while $\bm{f}^{\rho}=\rho (\boldsymbol{1}: \dot{\bm{\varepsilon}})$ is given by Eq. (\ref{eq:mass}).

\section{NEURAL INTEGRATION FOR CONSTITUTIVE EQUATIONS}
\label{sec:NDCE}
\noindent Building upon the integral formulation provided by neural differential equations (subsection \ref{par:neural}) and relying on the thermodynamic structure (subsection \ref{par:thermodynamics}), we devise a novel deep learning framework for the discovery of constitutive equations in small data regimes. The appeal of the approach is that we require neither (1) a dense time sampling of the state space $\mathcal{X}$ nor (2) a complete measurement of the data. Furthermore, the approach is applicable even if each variable of interest (state $\mathcal{X}$ and stress $\bm{\sigma}$) is acquired at irregular sampling times, in contrast with the common incremental formulation (subsection \ref{par:incremental}).

To achieve this goal, we aim at identifying the evolution equations, $\bm{f}\left(\mathcal{X},\dot{\bm{\varepsilon}}\right)$, and the internal energy function, $\widehat{u}\left(\mathcal{X} \right)$, in order to obtain thermodynamic admissible material models from data, \emph{cf.} Figure \ref{fig:intro}. In doing so, we replace the unknown functions with two neural operators, namely $\bm{f}_{\bm{\theta}}(\mathcal{X},\dot{\bm{\varepsilon}})$ and ${u}_{\bm{\omega}}(\mathcal{X})$, with parameters, respectively, $\bm{\theta}$ and $\bm{\omega}$. Thus, we rewrite the constitutive equations, according to Eq. (\ref{eqs:stress_dissipation}), as
\begin{subequations}
\label{eq:IVP_general}
\begin{align}
\label{eq:IVP_general_S}
{\mathcal{X}}_{{\bm{\theta}}}(t) &= \big \{ \rho_{\bm{\theta}}, \bm{\varepsilon}^e_{\bm{\theta}} , \mathbf{z}_{\bm{\theta}} \big \}(t) =  
\underset{\left(0, t\right]}{\mathrm{ODESolve}}\left( \mathcal{X}^{(0)}, \bm{f}_{{\bm{\theta}}}, \dot{\bm{\varepsilon}} \right), \qquad t \in \left[0,T\right]\\
\label{eq:IVP_general_stress}
{\bm{\sigma}}_{\bm{\theta}\bm{\omega}}(t) &= \dfrac{\partial {u}_{\bm{\omega}}}{\partial \bm{\varepsilon}^e_{\bm{\theta}}} + \left( \rho_{\bm{\theta}} \dfrac{\partial {u}_{\bm{\omega}}}{\partial \rho_{\bm{\theta}}} -{u}_{{\bm{\omega}}}\right)\boldsymbol{1}, \\
\label{eq:IVP_general_dissipation}
{d}_{\bm{\theta}\bm{\omega}}(t) &= \dfrac{\partial {u}_{\bm{\omega}}}{\partial \bm{\varepsilon}^e_{\bm{\theta}}} :\Big(\dot{\bm{\varepsilon}}-\bm{f}_{\bm{\theta}}^{\bm{\varepsilon}^e}\Big)-\dfrac{\partial {u}_{\bm{\omega}}}{\partial \mathbf{z}_{\bm{\theta}}}\cdot \bm{f}_{{\bm{\theta}}}^{\mathbf{z}} \geq 0,
\end{align}
\end{subequations}
where $\bm{f}_{\bm{\theta}} = \{ \bm{f}^{\rho}, \bm{f}_{\bm{\theta}}^{\bm{\varepsilon}^e},\bm{f}_{\bm{\theta}}^{\mathbf{z}}\}$.\\
%where $\mathcal{X}^{(0)}$ denotes the initial conditions for the state variables.

Equation (\ref{eq:IVP_general_S}) corresponds to the solution of the initial value problem related to the evolution equations, where $\mathcal{X}(0)=\mathcal{X}^{(0)}$ is the initial condition. The neural operator $\bm{f}_{\bm{\theta}}$ is treated as a neural differential equation, using the explicit midpoint method with fixed time step $h$ and global error of order $\mathcal{O}\left(h^2\right)$. Note that we focus here on a (total) strain control loading protocol, where the strain rates are provided at any given time. The primary goal in this case is to calculate the stress and state variables at each time point. In the evolution equation network, the strain rate thus acts as a control variable $\mathbf{q}\leftarrow \dot{\bm{\varepsilon}}$, \emph{cf.} Eq. (\ref{eq:neuralODE}). Equations (\ref{eq:IVP_general_stress}-\ref{eq:IVP_general_dissipation}) originate instead from the developments presented in subsection \ref{par:thermodynamics}, where $\widehat{u}$ has been replaced by its neural network approximation ${u}_{\bm{\omega}}$.

\subsection{Initial conditions}
\label{par:IC}
\noindent To predict the time evolution of the state variables, initial conditions $\mathcal{X}^{(0)} =\{\rho, \bm{\varepsilon}^e, \mathbf{z} \}^{(0)}$ are required. However, in contrast with the density and the dissipative state variables, the initial conditions for the elastic strain cannot be identified without either (\emph{i}) prescribing \textit{a priori} initial values (\emph{e.g.} zero) or (\emph{ii}) assuming a particular form of the internal energy and determining the material elastic moduli (using pressure and shear wave velocity measurements, for instance). Yet, these two paths carry intrinsic assumptions on the material behaviour and may prevent the identification of (thermodynamically admissible) constitutive equations that best fit the measurement data. 

With the aim of developing a scalable approach able to deal with the broadest range of materials and initial conditions, we propose to identify the initial conditions for the elastic strain from the knowledge of the initial values of the remaining state variables $\{\rho^{(0)}, \mathbf{z}^{(0)}\}$ and the stress $\bm{\sigma}^{(0)}$, by solving the following nonlinear equation
\begin{equation}
\text{find } \bm{\varepsilon}^{e(0)}  \text{ such that } \mathbf{r}^{(0)}_{\bm{\varepsilon}^{e}} = \mathbf{0}, \text{ with } \mathbf{r}^{(0)}_{\bm{\varepsilon}^{e}} \equiv  {\bm{\sigma}}_{\bm{\theta}\bm{\omega}}(0) -  \bm{\sigma}^{(0)}, 
\label{eq:initial_ee0}
\end{equation}
where ${\bm{\sigma}}_{\bm{\theta}\bm{\omega}}(0)$ is the thermodynamic expression of the material stress (\ref{eq:IVP_general_stress}), evaluated at time $t=0$, and $\bm{\sigma}^{(0)}$ is the corresponding measured value. For the solution of the nonlinear problem (\ref{eq:initial_ee0}), we treat $\bm{\varepsilon}^{e(0)}$ as a learnable hyper-parameter, which, initialised at zero, is iteratively updated using gradient descent by back-propagating the residual $\smash{\mathbf{r}^{(0)}_{\bm{\varepsilon}^{e}}}$ (see subsection \ref{subsec:learning}).

For the sake of clarity, it is worth noticing that the system of nonlinear equations in problem (\ref{eq:initial_ee0}) is underdetermined, due to the countless internal energy functions ${u}_{\bm{\omega}}$ for which the residual $\smash{\mathbf{r}^{(0)}_{\bm{\varepsilon}^{e}}}$ is zero, \emph{i.e.}, the number of parameters $\bm{\omega}$ is much larger than the number of components of the stress tensor. However, the proposed technique allows for the identification of only those values of the elastic strain that are thermodynamically admissible, \emph{i.e.}, in agreement with the Clausius-Duhem inequality, without the need of introducing additional biases.

\subsection{Learning process}
\label{subsec:learning}

\noindent Given the small data regime and in the absence of measurements of the elastic strain, the learning process of the neural integral constitutive equations, presented in Algorithm \ref{algo}, consists of the minimisation of the following loss function
\begin{equation}
\mathcal{L} =  \mathcal{L}_{\bm{\sigma}} +\mathcal{L}_{\mathbf{z}} + \mathcal{L}_{d}+\lVert \mathbf{r}^{(0)}_{\bm{\varepsilon}^{e}} \rVert_2^2+\lambda \Big(\lVert \bm{\theta} \rVert_2 +\lVert \bm{\omega} \rVert_2\Big),
\label{eq:loss}
\end{equation}
where $\mathcal{L}_{\bm{\sigma}}$ and $\mathcal{L}_{\mathbf{z}}$ are the supervised losses related to the predictions of the stress and dissipative state variables, respectively; $\mathcal{L}_{d}$ is an unsupervised loss for the fulfilment of the dissipation inequality (\ref{eq:IVP_general_dissipation}); $\mathbf{r}^{(0)}_{\bm{\varepsilon}^{e}}$ is the residual for the fulfilment of the initial conditions of the elastic strain, given by Eq. (\ref{eq:initial_ee0}); $\lVert\, \cdot \,\rVert_2 $ represents the $\ell_2$ norm; and $\lambda$ is a weighting parameter controlling the importance of the $\ell_2$ regularization (weight decay) of the parameters $\bm{\theta}$ and $\bm{\omega}$. Weight decay is here considered to penalise complexity of the neural networks and reduce eventual overfitting. We write
\begin{align}
\label{eq:loss_stress}
\mathcal{L}_{\bm{\sigma}} &= \dfrac{1}{N} \sum_{n = 1}^{N} \Big({\bm{\sigma}}_{\bm{\theta}\bm{\omega}}\left(t^{n} \right) - \bm{\sigma}^{(n)} \Big)^2,\\
\label{eq:loss_evol}
\mathcal{L}_{\mathbf{z}} &= \dfrac{1}{M} \sum_{m = 1}^{M}\bigg( \underset{\left(0, t^m\right]}{\mathrm{ODESolve}}\left( \mathbf{z}^{(0)}, \bm{f}^{\mathbf{z}}_{\bm{\theta}}, \dot{\bm{\varepsilon}} \right) -\mathbf{z}^{(m)} \bigg)^2,\\
\label{eq:loss_dissipation}
\mathcal{L}_{d} &= \dfrac{1}{N} \sum_{n = 1}^{N} \Bigl \langle -{d}_{\bm{\theta}\bm{\omega}}\left(t^{n}\right) \Bigl \rangle.
\end{align}
Here, $\{t^n,\bm{\sigma}^{(n)},\bm{\varepsilon}^{(n)}\}_{n=1}^{N}$ denote the $N$ sampling times of the stress measurements and protocol (control); $\{t^m,\mathbf{z}^{(m)}\}_{m=1}^{M}$ specify the $M$ sampling time of the acquisition of the state variables $\mathbf{z}$; and $\langle\,x \,\rangle \equiv \left(|x|+x\right)/2$ are the Macaulay brackets. Note that the quantities involved in Eqs. (\ref{eq:loss}-\ref{eq:loss_dissipation}) are all dimensionless and normalised in the range $[-1,1]$, except for the time scale which is normalised in the range $[0,1]$. For more detail, we refer to the repository accompanying this manuscript \cite{nice}.

The distinction between $N$ and $M$ allows one to operate with arbitrary sampling frequencies of the measurement data. Note that $M \ll N$ in small data regimes, and that $t^N\equiv t^M \equiv T$, where $T$ represents the last time point, $t\in [0,T]$. The error related to the dissipation inequality is evaluated at the same sampling times of the stress measurements, $t^n$.

We use early stopping criterion based on the loss (\ref{eq:loss}) associated with the validation data set to stop the training process and avoid overfitting (\emph{cf.} Section \ref{sec:benchmarks}).

\begin{algorithm}
  \caption{Pseudocode of the neural integration for constitutive equations.}
  \label{algo}
\small %\sf
  \SetKwInOut{Require}{Require}
  \Require{%
    network params $\{\bm{\theta}, \bm{\omega}\}$; hyper-params $\bm{\varepsilon}^{e(0)}$; optimizer.
  }\vspace{1.5pt}
  \KwData{initial conditions $\{\rho, \bm{\sigma}, \mathbf{z}\}^{(0)}$; data and protocol $\{t^n, \bm{\sigma}^{(n)}, \bm{\varepsilon}^{(n)}\}_{n=1}^{N}$, $\{t^m, \mathbf{z}^{(m)}\}_{m=1}^{M}$.}
\vspace{1.5pt}
  \While{training}{
    $\mathcal{X}^{(0)} \leftarrow \{\rho^{(0)}, \bm{\varepsilon}^{e(0)}, \mathbf{z}^{(0)}\}$\;
    compute ${\mathcal{X}}_{{\bm{\theta}}}(t) = \underset{\left[0, t\right]}{\mathrm{ODESolve}}\big( \mathcal{X}^{(0)}, \bm{f}_{{\bm{\theta}}}, \dot{\bm{\varepsilon}} \big)$\;
    
    \For{$t \gets 0$ \KwTo $T$ }{
      compute ${u}_{\bm{\omega}}\left(\rho, \bm{\varepsilon}^e, \mathbf{z}\right)$\;
      compute $\frac{\partial {u}_{\bm{\omega}}}{\partial \mathcal{X}_{\bm{\theta}}}$ and evaluate ${\bm{\sigma}}_{\bm{\theta}\bm{\omega}}$, ${d}_{\bm{\theta}\bm{\omega}}$
      %\tcp*{autodiff using Eqs.(\ref{eq:IVP_general_stress}-\ref{eq:IVP_general_dissipation})}
    }
    
    compute and backpropagate loss $\mathcal{L}$ at times $\{t^n\}_{n=1}^{N}$, $\{t^m\}_{m=1}^{M}$\;
    %\tcp*{adjoint sensitivity}
    update $\{\bm{\theta}, \bm{\omega}, \bm{\varepsilon}^{e(0)}\}$\;
    
    \If{early-stopping criterion}{
      \emph{training} false\;
    }
  }  
  \Return{params $\{\bm{\theta}, \bm{\omega}\}$}

\end{algorithm}

It is also worth mentioning that, even in the absence of the entire time evolution of the elastic strain, only the initial values of the latter have to be identified. Indeed, the time integral formulation of the evolution equations allows learning the trajectory of $\bm{\varepsilon}^e$ by minimisation of the error related to the material stress. This is an important property of the approach proposed as it does not only enables learning from limited (sparse) data but also in conditions where only partial information about the state of the material is available. 

Finally, notice that the loss function (\ref{eq:loss}) does not include errors related to the prediction of the values of the specific internal energy and the dissipation rate (in contrast, for instance, with \citep{masi2023evolution}), as the latter are not necessary conditions for the fulfilment of the laws of thermodynamics and the accurate prediction of the material response. 

\section{NUMERICAL BENCHMARKS}
\label{sec:benchmarks}
\noindent The accuracy and capability of the developed approach are evaluated using synthetic data generated using analytical constitutive models. Although testing with real experimental data is the ultimate goal, using numerically generated data offers the benefit of having exact knowledge of the ground truth, \emph{i.e.}, the constitutive model that underlies the data.

To mimic idealised experiments, we generate numerical data sets for training, validation, and testing using standard protocols in experimental mechanics, which are characterised by monotonous loading of the specimen and unloading to an unstressed configuration. In so doing, we disregard cyclic loading paths and other (more) exotic protocols in the generation of data for training the model, as those are rarely carried out in reality. For each protocol, we sample the stress values at $1/N$ time points and the dissipative state variables at $1/M$ time points (\ref{eq:IVP_general}). In addition, we assess the robustness of the approach with respect to errors that may be present in the data used in the training process. Such errors are simulated by generating synthetic noise, sampled from a normal distribution and added to the ground-truth data sets.

We select two benchmarks that highlight the challenges in learning constitutive representations in small data regimes: a one-state variable model ($\bm{\varepsilon}^e$) and a three-state variables model ($\rho,\bm{\varepsilon}^e,\mathbf{z}$). 

In all cases, the entire data set is split into training, validation, and test sets, following approximately the following proportions: 60-20-20\%.  The new algorithm is implemented using the deep learning library \texttt{PyTorch} \citep{paszke2017automatic} and the package \texttt{torchdiffeq} \citep{torchdiffeq} for the time integration of the evolution equations and the back-propagation framework to determine gradients. The time step of the explicit midpoint method is selected as $h= \frac{T}{ 800}$.

For the sake of conciseness, the comparison of the proposed approach with the common incremental formulation is discussed in \ref{appendixA}.

\subsection{Elasto-plastic media}
\label{subsec:DPu}
\noindent The first example concerns an incrementally nonlinear elasto-plastic, non-associative, Drucker-Prager material, with constant density (\emph{cf.} Sect. 8 in \citep{einav2012unification}). The specific internal energy is defined as
\begin{equation}
\widehat{u}(\bm{\varepsilon^e}) = \dfrac{1}{2}K{\varepsilon_v^e}^2+\dfrac{3}{2}G{\varepsilon^e_s}^2,
\label{eq:energy_DP}
\end{equation}
where $\varepsilon_v^e=\boldsymbol{1}: \bm{\varepsilon}^e$ is the volumetric invariant of the strain tensor and ${\varepsilon_s^e = \sqrt{\sfrac{2}{3}\;{\bm{\varepsilon}^e}^{\prime}\hspace{-3pt}:{\bm{\varepsilon}^e}^{\prime}}}$ is the invariant of the deviatoric strain tensor,  where ${\bm{\varepsilon}^e}^{\prime} = \bm{\varepsilon}^e-\frac{\varepsilon^e_v}{3}\boldsymbol{1}$. We select $K=70$ GPa, $G=60$ GPa, friction coefficient equal to one, seismic parameter equal to one, and dilatancy parameter equal to one half (\emph{cf.} \citep{einav2012unification}).

Note that the entire material response is described by a single state variable ($\bm{\varepsilon}^e$), thus the loss term $\mathcal{L}_{\mathbf{z}}$ is neglected. Despite the simplicity of such material model, this example serves as a proof of principle of our approach in retrieving evolution equations and material response from partial measurement data, with and without synthetic noise. 

\subsubsection{Data generation}
\noindent We generate data by applying single loading-unloading paths, with constant strain rate, and integrating in time the evolution equation for the elastic strain rate (using Eq. 8.15 in \citep{einav2012unification}) to compute the evolution of stress according to the energy function (\ref{eq:energy_DP}). The material stress is here identified in terms of its volumetric and deviatoric parts: $p=\boldsymbol{1}:\frac{\bm{\sigma}}{3}$ and $\smash{q = \sqrt{{\sfrac{3}{2}}\;\bm{\sigma}^{\prime}\hspace{-3pt}:\bm{\sigma}^{\prime}}}$, with $\bm{\sigma}^{\prime} = \bm{\sigma}-p\boldsymbol{1}$.

\begin{figure}[h]
\centering
\begin{subfigure}{0.34\textwidth}
\centering
\includegraphics[width=0.9\textwidth]{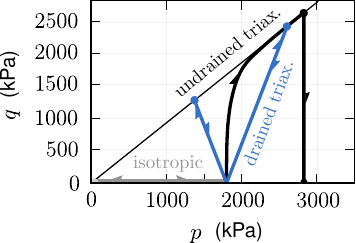}
\caption{\centering loading paths and protocols}
\label{fig:training_data_loss_a}
\end{subfigure}
\hspace{10pt}
\begin{subfigure}{0.44\textwidth}
\centering
\vspace{12pt}
\includegraphics[width=0.9\textwidth]{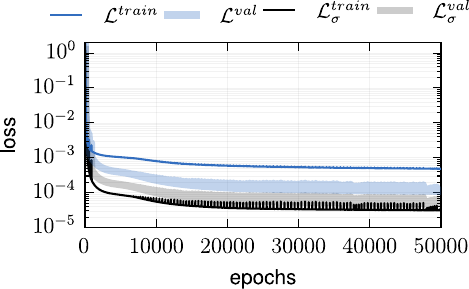}
\caption{\centering training and validation losses}
\label{fig:training_data_loss_b}
\end{subfigure}
\caption{Test cases for training and validation. (a) Loading protocols for the generated data and (b) the evolution of training and validation losses during training. The generated data sets involve drained triaxial ($\dot{p} = \pm \dot{q}/3$), undrained triaxial ($\dot{\varepsilon}_v=0$), and isotropic compression ($\dot{\varepsilon}_s=0$), where the narrow line represents critical state.}
\label{fig:training_data_loss}
\end{figure}

From an initial unstressed configuration, the material is first subjected to isotropic compression to generate states at different confining pressures, $p\in [200, 2200]$ kPa. This then represents the initial configuration for subsequent numerical experiments composed of the following loading paths: isotropic compression/extension, undrained compression, and drained triaxial compression/extension (see Figure \ref{fig:training_data_loss_a}). Each path is composed of $N=40$ measurements of the stress values.

The neural operator $\bm{f}_{\bm{\theta}}$ is composed of three hidden layers with 36 nodes (Gaussian error linear unit activations, $\mathrm{GELU}\left(\sq\right) = 1.7\sq^2/(1+e^{-1.7\sq })$) and one linear output layer, while ${u}_{\bm{\omega}}$ consists of two hidden layers with  64 units (softplus activations $\mathrm{s}^+\left(\sq\right) = \log(1+e^{\sq} )$) and one linear output layer. The training is performed for the entire batch of training data, using Adam optimiser, with adaptive learning rate, with upper and lower bounds equal to 10\textsuperscript{-2} and 10\textsuperscript{-4}, respectively.

\subsubsection{Noise-free measurement data} 
\noindent At first, we consider data free of any noise. Figure \ref{fig:training_data_loss_b} displays the evolution of the training and validation losses during the learning process. Convergence is reached after approximately 20'000 epochs with a mean absolute percentage error in the stress prediction for the test set equal to 0.9\%.

\begin{figure}[h]
\centering
\includegraphics[width=0.8\textwidth]{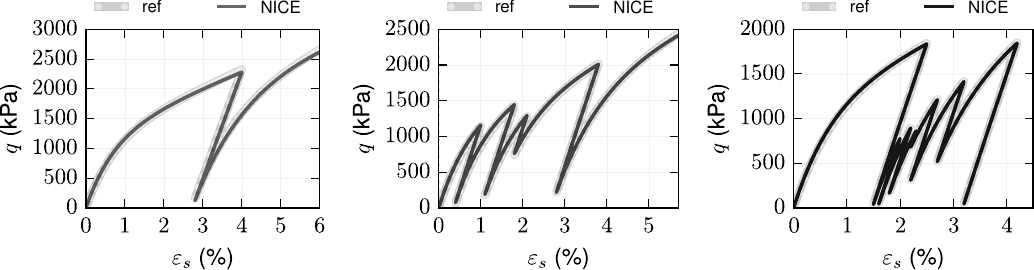}
\caption{Predictions at inference for an undrained cyclic loading path (unobserved protocol) with one, four, and six unloading-reloading cycles (from left to right).} 
\label{fig:DP_freq_UT}
\end{figure}

\begin{figure}[h]
\centering
\includegraphics[width=0.8\textwidth]{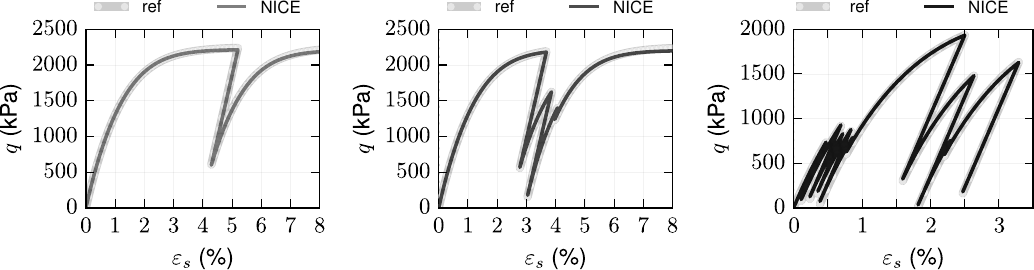}
\caption{Predictions at inference for a drained triaxial cyclic loading path (unobserved protocol) with one, three, and nine unloading-reloading cycles (from left to right).}
\label{fig:DP_freq_DT}
\end{figure}

The model is then deployed to infer the stress-strain behaviour for new, unobserved loading protocols and paths. Figures \ref{fig:DP_freq_UT} and \ref{fig:DP_freq_DT} compare the predictions (denoted with the label \textsf{NICE}) with the ground-truth results (denoted by \textsf{ref}) for cyclical undrained and drained triaxial loading paths, respectively. Notice that the neural model enables the prediction of consecutive unloading and reloading sequences, despite the learning process concerned only simple monotonous protocols.  Despite some minor differences, the approach yields accurate predictions with respect to the reference values, independently of the particular loading protocol.

\begin{figure}[h]
\centering
\includegraphics[width=0.8\textwidth]{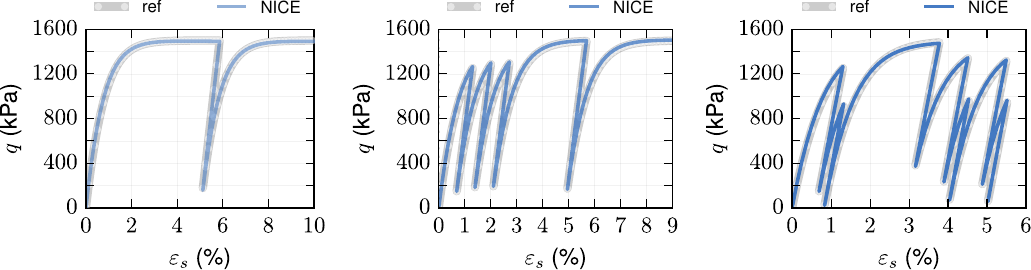}
\caption{Predictions at inference for an unobserved cyclical loading path (drained triaxial at constant pressure, $\dot{p}=0$ and axial strain rate prescribed), with unobserved protocol characterised by one, three, and nine unloading-reloading cycles (from left to right).}
\label{fig:DP_freq_DTP}
\end{figure}

The neural integration for constitutive equations additionally allows modelling unobserved loading paths -- that is, predictions under strain configurations that do not belong to the training and validation sets. This is presented in Figure \ref{fig:DP_freq_DTP}, where the predictions are shown for a drained triaxial path at constant pressure, \emph{i.e.}, $\dot{p}=0$, and prescribed axial strain rate, and unseen cyclical loading protocol. Once more, the accuracy of the predictions is independent of the frequency and number of the unloading-reloading cycles.

\subsubsection{Influence of noise in the training and validation data sets} 
\label{par:noise_DP}
\begin{figure}[h]
\centering
\includegraphics[width=0.9\textwidth]{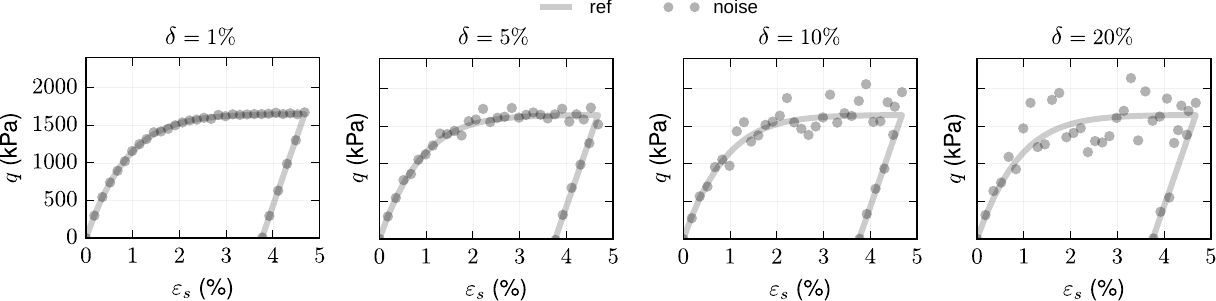}
\caption{Effect of the synthetic noise amplitude ($\delta$) in terms of the stress-strain response for one of the loading-unloading paths used for training.}
\label{fig:pred_noiseDP_training}
\end{figure}

\noindent To mimic noise in an hypothetical data acquisition system, the original training and validation data sets are corrupted with synthetic noise according to
\begin{equation}
\mathbf{x}^* = \mathbf{x}+{\mu}_{\mathbf{x}}\, \mathcal{N}\left(\mathbf{0},\delta \boldsymbol{1}\right),
\label{eq:noise}
\end{equation}
where $\mathbf{x}$ represents the ground-truth values of a generic quantity in terms of the stress and the state variables; $\mathbf{x}^*$ are the values corrupted with noise; $\mathcal{N}\left(\mathbf{0},\delta\right)$ is a normal distribution with zero mean and standard deviation equal to $\delta$; ${\mu}_{\mathbf{x}}$ is the mean value of $\mathbf{x}$; and $\delta$ is the noise percentage. Figure \ref{fig:pred_noiseDP_training} displays the influence of the amplitude of the synthetic noise on the data used for training in terms of the stress-strain response for one of the loading-unloading paths.

\begin{figure}[th]
\centering
\includegraphics[width=0.45\textwidth]{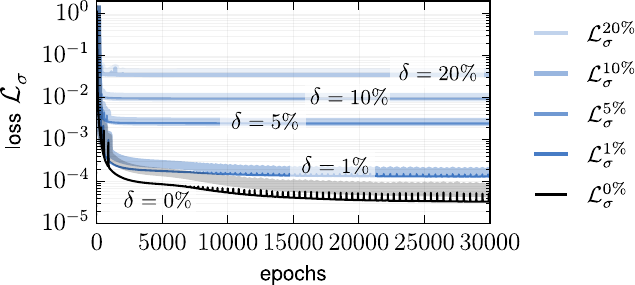}
\caption{Evolution of training (unshaded) and validation (shaded) loss during training, at varying noise amplitudes used to corrupt the training data.}
\label{fig:loss_noiseDP}
\end{figure}

The architectures of the neural network and the training strategy are kept the same as in the previous scenario. Figure \ref{fig:loss_noiseDP} displays the evolution of the training loss $\mathcal{L}_{\bm{\sigma}}$ for the different noise levels. Notice that the final value of the loss depends on the amount of synthetic noise suggesting that the network does not overfit the corrupted data (\emph{cf.} \citep{masi2021thermodynamics}).

\begin{figure}[h]
\centering
\begin{subfigure}{0.3\textwidth}
\centering
\includegraphics[height=0.3\textheight]{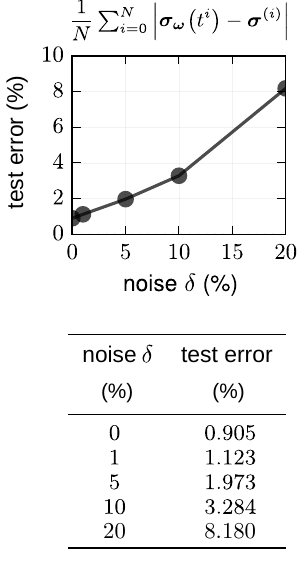}
\caption{\centering test: mean absolute error}
\label{fig:pred_noiseDPa}
\end{subfigure}
\begin{subfigure}{0.66\textwidth}
\centering
\includegraphics[height=0.3\textheight]{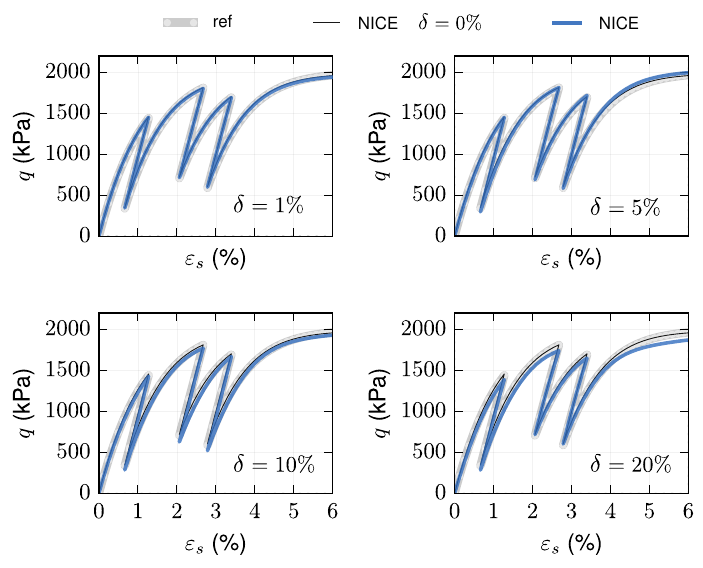}
\caption{\centering inference: stress predictions}
\label{fig:pred_noiseDPb}
\end{subfigure}
\caption{Performance and robustness of the approach at varying amplitudes of the synthetic noise ($\delta$) added to the training and validation sets.  (a) Mean absolute percentage error of the network predictions with respect to the un-noisy test set, and (b) predictions for an unseen constant pressure cyclical drained triaxial loading path for $\delta=1,5\%$ (top) and $\delta=10,20\%$ (bottom).}
\label{fig:pred_noiseDP}
\end{figure}

Given the noise, the predictions by the proposed neural framework are tested for unseen loading conditions. Figure \ref{fig:pred_noiseDPa} presents the mean absolute percentage error of the stress predictions for the un-noisy test set for different levels of synthetic noise ($\delta$). Figure \ref{fig:pred_noiseDPb} shows the predictions of the network trained on corrupted data sets with those obtained without synthetic noise and the ground-truth solutions, for unobserved loading-unloading cycles.

The newly proposed approach enables high degrees of robustness with respect to noisy data. In particular, the error of the predictions for the test set is always smaller than the mean value of the synthetic noise added to the training data (8\% versus 20\%). This is thanks, in part to the hardwiring of the existence of the energy potential and entropy balance in the formulation (as demonstrated in \citep{masi2021thermodynamics}) but primarily thanks to the integral formulation. Indeed, the latter is intrinsically robust to high noise amplitudes (\emph{cf.} gated integration, \citep{ware1966high}), especially when compared with the conventional incremental formulation (see \ref{appendixA}).

\subsection{Elasto-plastic porous media}
\label{subsec:MCBu}

\noindent The second example concerns elasto-plastic porous media (see \citep{riley2023constitutive}, for more). The reference constitutive model is built on three state variables: elastic strain ($\varepsilon^e$), bulk density ($\rho$), and solid fraction $\phi=\rho/\rho_s$, where $\rho_s$ is the solid density, defined as the ratio of the mass of the solid phase over its volume. The same model can additionally be extended to brittle materials by accounting for the dependence on the granular temperature, although this is neglected in the present work (by selecting $c=0$, Eq. 19 in \citep{riley2023constitutive}).

Following this model, the specific internal energy is defined as
\begin{equation}
\widehat{u}(\rho,\bm{\varepsilon^e}) = \left( \dfrac{\rho}{\rho_s^*}\right) \left( \dfrac{1}{2}K{\varepsilon_v^e}^2+\dfrac{3}{2}G{\varepsilon^e_s}^2\right),
\label{eq:energy_DR}
\end{equation}
where $\rho_s^*$ is the unstressed solid density, which is a material parameter. Despite the aforementioned internal energy function does not depend on the solid fraction, the latter is responsible for density hardening of the bounding surface. The evaluation is based on the following model parameters: bulk modulus $K=10$ MPa and shear modulus $G=6$ MPa, the slope of the critical state line $M=1.5$, the dimensionless parameter that determines the effective isotropic yield pressure $\beta^*=0.1$, and the unstressed solid density $\rho_s^*=600$ kg/m\textsuperscript{3}. For more, we refer to \cite{riley2023constitutive}.

In deploying the proposed approach, the internal energy network is considered with the classification provided by Eq. (\ref{eq:energy}). Unlike the previous case, here the model specifically includes $\mathbf{z} \equiv \phi$ by virtue of the axiom of equipresence (see \citep{truesdell1960classical}).
In contrast with the first benchmark, we deploy the complete formulation of the neural integral constitutive equations (presented in Section \ref{sec:NDCE}), using the loss function given by Eq. (\ref{eq:loss}), and showcase the capabilities of the neural differential approach in presence of scarce and incomplete data.

\subsubsection{Data generation}
\noindent Data are generated by applying single loading-unloading strain-loading paths, with constant strain rate, as for the previous benchmark. In order to characterise the influence of the solid fraction, four specimens are considered with $\phi^{\mathrm{init}} = {0.5,0.6,0.7,0.8}$ at the initial unstressed configuration. This is done to mimic a hypothetical experimental acquisition where the state variables of only a finite and reduced number of configurations can be actually investigated.
\begin{figure}[h]
\centering
\begin{subfigure}{0.25\textwidth}
\centering
\includegraphics[width=0.9\textwidth]{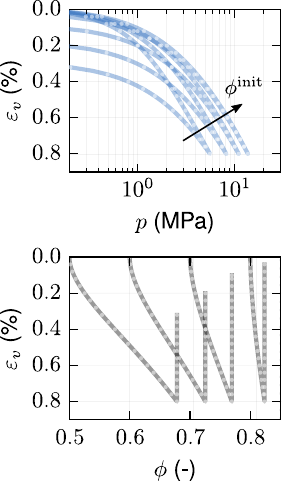}
\caption{\centering isotropic compression}
\label{fig:path_REa}
\end{subfigure}
\hspace{10pt}
\begin{subfigure}{0.52\textwidth}
\centering
\includegraphics[width=0.9\textwidth]{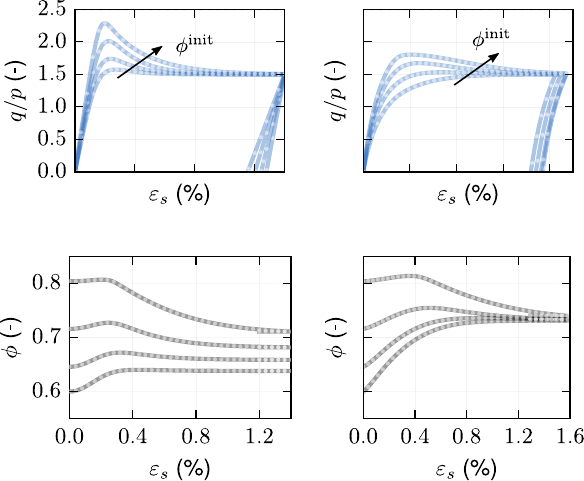}
\caption{\centering undrained (left) and drained (right) triaxial}
\label{fig:path_REb}
\end{subfigure}
\caption{Loading paths and protocols for training, validation, and test data sets. (a) Isotropic compression loading tests followed by extensional unloading with four initial values of the solid fractions ($\phi^{\mathrm{init}}$), and (b) undrained and drained loading-unloading triaxial compression for an initial confining pressure $p^{(0)}=1.0$ MPa, starting from four different solid fractions.}
\label{fig:path_RE}
\end{figure}

From the unstressed configuration, we apply isotropic compression to generate states at different confining pressures (see Figure \ref{fig:path_REa}), $p\in [1000, 8500]$ kPa, which are followed by numerical experiments composed of undrained and drained triaxial compression and isotropic extension (see Figure \ref{fig:path_REb}).

Each path is composed of $N=140$ measurements of the stress. To probe the ability of our approach in learning from small data, two acquisition strategies are considered for the dissipative state variable values: $M=N$ and $M=2$. The former represents a dense time acquisition of the material state, while the latter represents the frequent scenario depicted in Figure \ref{fig:intro}, where only the initial and final values of the material state are known thanks to measurements. Note that in both scenarios, as in reality, the observations of the elastic strain are unavailable.

The neural operator $\bm{f}_{\bm{\theta}}$ is composed of three hidden layers with 42 nodes (Gaussian error linear unit activations) and one linear output layer, while ${u}_{\bm{\omega}}$ consists of two hidden layers with 64 nodes (softplus activations) and a linear output layer. The same training procedure detailed in subsection \ref{subsec:DPu} is followed.

\subsubsection{Noise-free measurement data} 
\begin{figure}[th!]
\centering
\begin{subfigure}{0.287\textwidth}
\centering
\includegraphics[width=0.8\linewidth]{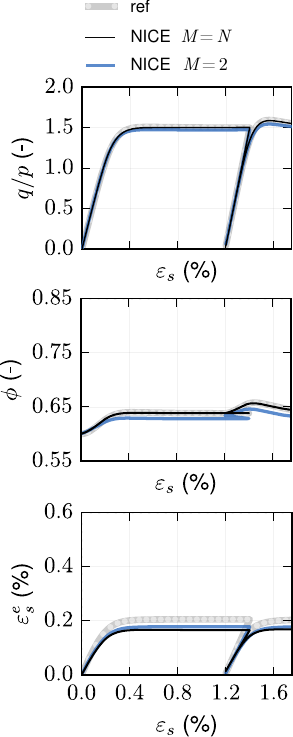}
\caption{\centering $p^{(0)} = 2500$ kPa}
\label{fig:pred_RE_MN2a}
\end{subfigure}
\hspace{2pt}
\begin{subfigure}{0.245\textwidth}
\centering
\includegraphics[width=0.8\linewidth]{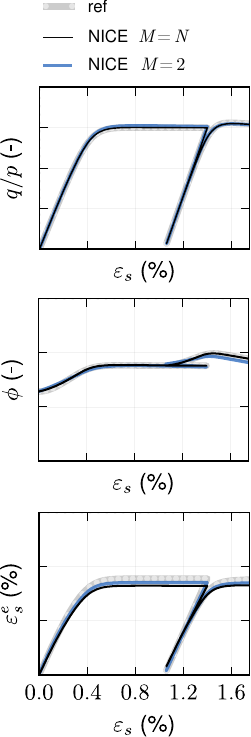}
\caption{\centering $p^{(0)} = 5000$ kPa}
\label{fig:pred_RE_MN2b}
\end{subfigure}
\hspace{2pt}
\begin{subfigure}{0.245\textwidth}
\centering
\includegraphics[width=0.8\linewidth]{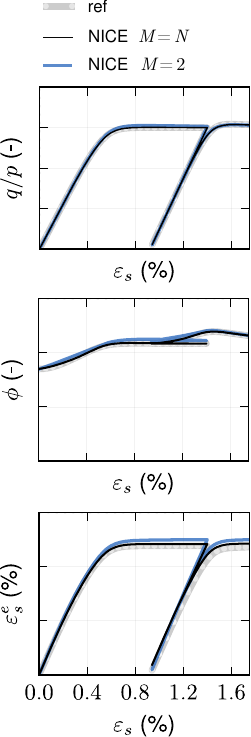}
\caption{\centering $p^{(0)} = 8500$ kPa}
\label{fig:pred_RE_MN2c}
\end{subfigure}
\caption{Predictions at inference in terms of the stress ratio ($q/p$, top), solid fraction ($\phi$, middle), and deviatoric elastic strain ($\varepsilon^e_s$, bottom) for an undrained triaxial compression loading-unloading-reloading cycles with unobserved loading protocol for the two configurations of the acquisition of the state variables: dense ($M=N$) and sparse ($M=2$). Material prepared with initial solid density $\phi^{\mathrm{init}}=0.5$ and subjected to isotropic compression up to different initial confining pressures $p^{(0)}=2500, 5500, 8500$ kPa, from left to right.}
\label{fig:pred_RE_MN2}
\end{figure}

\noindent First consider the idealised case of noise-free data sets for the training and validation of the deep learning algorithm. In particular, it is instructive to compare the accuracy and performance of the proposed formulation depending on the sampling frequency of the dissipative state variable, $\phi$. Thus, after training both configurations (\emph{i.e.}, $M=N$ and $M=2$), the predictions for the unobserved cyclical loading protocols are compared with a set of undrained triaxial compressions (Figure \ref{fig:pred_RE_MN2}).
Specifically, Figures \ref{fig:pred_RE_MN2a}, \ref{fig:pred_RE_MN2b}, and \ref{fig:pred_RE_MN2c} depict the constitutive response of specimens prepared from an initial unstressed configuration, with solid fractions of $\phi^{\mathrm{init}}=0.5$, and isotropically compressed upon reaching three different initial confining pressures, $p^{(0)}=2500, 5500, 8500$ kPa, respectively. The results are shown in terms of the stress ratio, solid fraction, and deviatoric elastic strain.

Regardless of the acquisition frequency, the proposed approach enables accurate constitutive representations of a material with multiple state variables from partial data. The method allows, as for the case of a simple elasto-plastic material, the modelling of cycles of unloading and reloading stages even in the absence of such protocols in the training data set. Despite the lack of any measurement of one state variable ($\bm{\varepsilon}^e$), both configurations accurately predict the evolution of the stresses and state variables, including the elastic strain. Note, however, that the discovered elastic strain coincides with the ground truth up to a constant term (which we removed in the plots in Figure \ref{fig:pred_RE_MN2}). Furthermore, a dense parametrisation of the solid fraction ($\mathbf{z}$) enables very accurate predictions in terms of all quantities of interest. But, the model continues to deliver high accuracies in the sparse acquisition configuration ($M=2$). This demonstrates the capabilities of the neural integral constitutive equations in learning constitutive models from small data regimes.

One minor difference exist between the two cases, $M=N$ and $M=2$. Relying on a dense parametrisation of $\phi$, the model with $M=N$ accurately identifies the plateau in the evolution of the solid fraction at unloading, while in the case $M=2$, for some loading paths, the solid fraction does not necessarily remain constant. The reason lies in the fact that the neural model with $M=2$ has not observed such a phenomenon through the given training data, and thus cannot fully grasp the steady evolution at unloading.\\ %Second, we can observe that the sparse neural model identifies slightly different plateaus, at relatively large deformations, in the case of the elastic strain (and also of the stress, when small confining pressures are considered).\\

\begin{figure}[h!]
\centering
\begin{subfigure}{0.255\textwidth}
\centering
\includegraphics[width=0.9\textwidth]{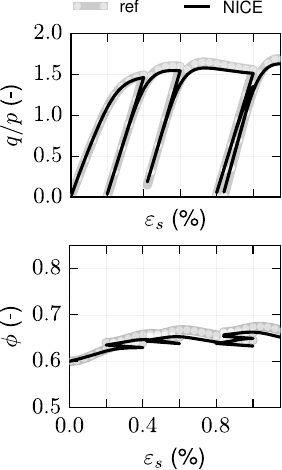}
\caption{\centering $p^{(0)}=2500$ kPa}
\label{fig:pred_RE_UTa}
\end{subfigure}
\hspace{1pt}
\begin{subfigure}{0.227\linewidth}
\centering
\includegraphics[width=0.9\textwidth]{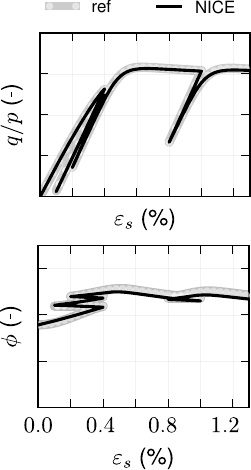}
\caption{\centering $p^{(0)}=5500$ kPa}
\label{fig:pred_RE_UTb}
\end{subfigure}
\hspace{1pt}
\begin{subfigure}{0.225\textwidth}
\centering
\includegraphics[width=0.9\textwidth]{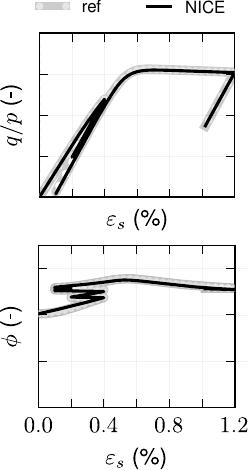}
\caption{\centering $p^{(0)}=7000$ kPa}
\label{fig:pred_RE_UTc}
\end{subfigure}
\hspace{1pt}
\begin{subfigure}{0.227\textwidth}
\centering
\includegraphics[width=0.9\textwidth]{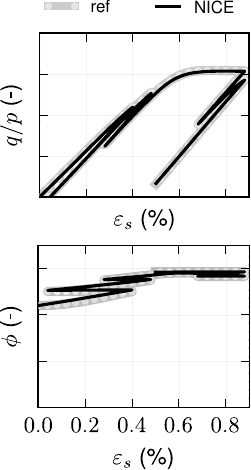}
\caption{\centering $p^{(0)}=8500$ kPa}
\label{fig:pred_RE_UTd}
\end{subfigure}
\caption{Predictions at inference ($M=2$) in terms of the stress ratio ($q/p$, top) and solid fraction ($\phi$, bottom) for an undrained triaxial compression with random cyclical loading protocols. Tests starts with a similar initial solid fraction $\phi^{\mathrm{init}}=0.5$ at the unstressed configuration and follows an isotropic compression to reach three different initial confining pressures, $p^{(0)}=2500, 5500, 7000, 8500$ kPa, from left to right.}
\label{fig:pred_RE_UT}
\end{figure}

\begin{figure}[h!]
\centering
\begin{subfigure}{0.262\textwidth}
\centering
\includegraphics[width=0.9\textwidth]{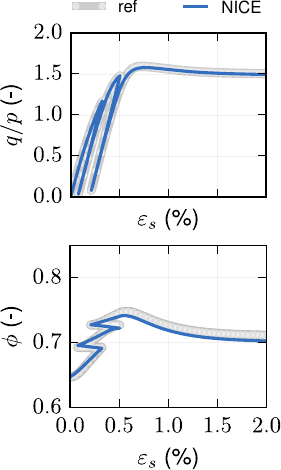}
\caption{\centering $\phi^{\mathrm{init}}=0.5$}
\label{fig:pred_RE_DTa}
\end{subfigure}
\begin{subfigure}{0.232\textwidth}
\centering
\includegraphics[width=0.9\textwidth]{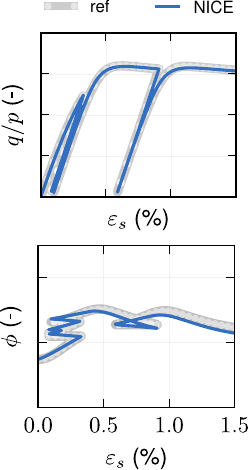}
\caption{\centering $\phi^{\mathrm{init}}=0.6$}
\label{fig:pred_RE_DTb}
\end{subfigure}
\begin{subfigure}{0.232\textwidth}
\centering
\includegraphics[width=0.9\textwidth]{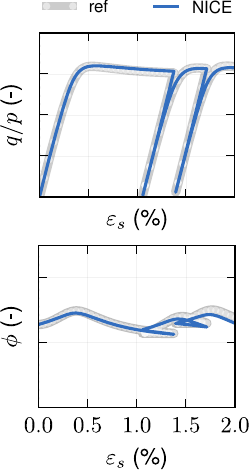}
\caption{\centering $\phi^{\mathrm{init}}=0.7$}
\label{fig:pred_RE_DTc}
\end{subfigure}
\begin{subfigure}{0.233\textwidth}
\centering
\includegraphics[width=0.9\textwidth]{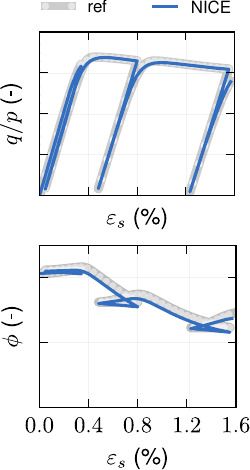}
\caption{\centering $\phi^{\mathrm{init}}=0.8$}
\label{fig:pred_RE_DTd}
\end{subfigure}
\caption{Predictions at inference ($M=2$) in terms of the stress ratio ($q/p$, top) and solid fraction ($\phi$, bottom) for an unobserved loading path (random loading protocols of drained cyclical triaxial tests at constant pressure, $\dot{p}=0$ with prescribed axial strain rate). In all the above the initial confining pressure is given by $p^{(0)}=2500$ kPa. However, the initial solid fraction of the unstressed configuration varies $\phi^{\mathrm{init}}=0.5, 0.6, 0.7, 0.8$, from left to right.}
\label{fig:pred_RE_DT}
\end{figure}

Next, the focus is directed towards probing the inference capabilities of the sparse neural model ($M=2$) under more complex loading scenarios. To this end, Figures \ref{fig:pred_RE_UT} and \ref{fig:pred_RE_DT} compare the approach with the reference constitutive model for a randomly cycled undrained triaxial and a drained triaxial tests path at constant pressure, respectively. The former consists of a specimen with initial solid fraction $\phi^{\mathrm{init}}=0.5$ and tests the predictions at varying of the initial confining pressure, $p^{(0)}$. The latter tests the influence of the initial solid fraction, for the same value of $p^{(0)}$. Notice that the model has not observed the drained triaxial test case. 


In both scenarios, the network predictions are in good agreement with the reference model. This is particularly true for the evolution of the stress, despite the demanding loading protocols involving many cycles with random unloading and reloading frequencies. Regarding the solid fraction, the network yields overall accurate predictions, despite some minor differences in the evolution during unloading, as mentioned above.

\begin{figure}[h]
\centering
\begin{subfigure}{0.27\textwidth}
\centering
\includegraphics[width=0.9\textwidth]{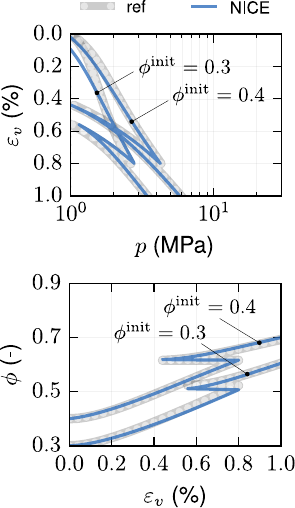}
\caption{\centering $\phi^{\mathrm{init}}=0.3,0.4$}
\label{fig:pred_RE_ISOa}
\end{subfigure}
\hspace{2pt}
\begin{subfigure}{0.25\textwidth}
\centering
\includegraphics[width=0.9\textwidth]{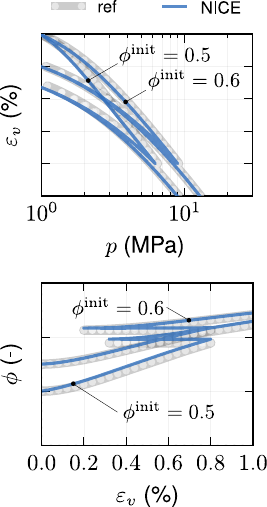}
\caption{\centering $\phi^{\mathrm{init}}=0.5,0.6$}
\label{fig:pred_RE_ISOb}
\end{subfigure}
\hspace{2pt}
\begin{subfigure}{0.25\textwidth}
\centering
\includegraphics[width=0.9\textwidth]{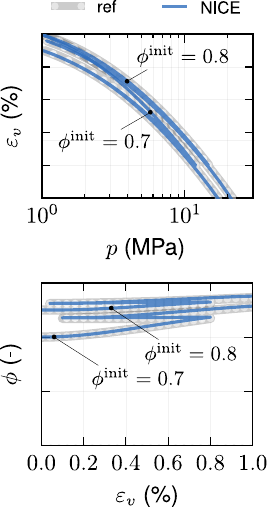}
\caption{\centering $\phi^{\mathrm{init}}=0.7,0.8$}
\label{fig:pred_RE_ISOc}
\end{subfigure}
\caption{Predictions at inference ($M=2$) in terms of the volumetric stress ($p$, top) and solid fraction ($\phi$, bottom) for isotropic compression with unobserved loading protocol at unobserved initial solid fractions: $\phi^{\mathrm{init}}=0.3\div0.8$, from left to right.}
\label{fig:pred_RE_ISO}
\end{figure}

As a final application, consider the case of cyclic isotropic compression-extension loading, from unstressed configurations, with unobserved initial values of the solid fraction, namely $\phi^{\mathrm{init}}=0.3\div 0.8$, as shown in Figure \ref{fig:pred_RE_ISO}. Despite some minor discrepancies for very small values of the solid fraction (Figure \ref{fig:pred_RE_ISOa}), the neural model yields good extrapolation capabilities, outside the range of values originally observed during training.

For the sake of completeness, the comparison with the incremental formulation is given in \ref{appendixA2}. There, we demonstrate that the conventional incremental data-fitting approach is incapable of discovering accurate constitutive models when faced with incomplete or sparse data sets. This holds true regardless of the thermodynamic structure hardwired in the neural network architecture, demonstrating the importance of the newly proposed integral formulation.

\subsubsection{Influence of noise in the training and validation data sets} 
\noindent To evaluate the robustness of the approach, the original training and validation sets are corrupted with synthetic noise (\emph{cf.} paragraph \ref{par:noise_DP}), while the training procedure is repeated using the neural network architecture discussed in subsection \ref{subsec:MCBu}. Note that noise is considered for all state variables, except the bulk density.
\begin{figure}[h]
\centering
\begin{subfigure}{0.49\textwidth}
\centering
\includegraphics[width=0.9\textwidth]{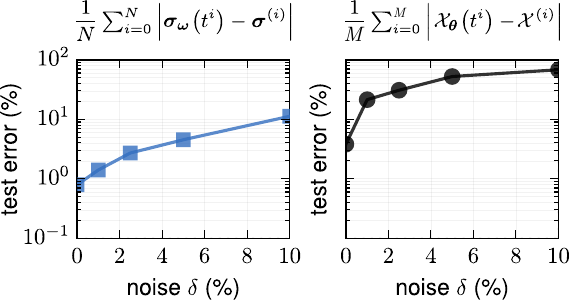}
\caption{\centering $M=N$}
\label{fig:noise_influence_REa}
\end{subfigure}
\begin{subfigure}{0.49\textwidth}
\centering
\includegraphics[width=0.9\textwidth]{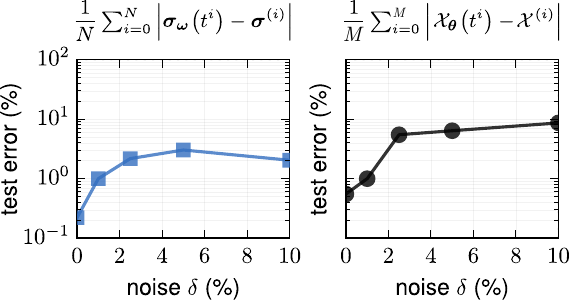}
\caption{\centering $M=2$}
\label{fig:noise_influence_REb}
\end{subfigure}
\caption{Mean absolute percentage error of the network predictions in terms of the stress ($\sigma$) and solid fraction ($\mathcal{X} = \phi$, dissipative state variable) with respect to the un-noisy test set for: (a) dense and (b) sparse sampling cases of the state variables.}
\label{fig:noise_influence_RE}
\end{figure}

In Figure \ref{fig:noise_influence_RE}, the mean absolute percentage error of the stress and solid fraction predictions for the noise-free test set are studied for varying degrees of the level $\delta$ of the synthetic noise. As before, we consider two acquisition configurations: one with $M=N$ and another with $M=2$.
For stress predictions, the error is consistently found to always be smaller or equal to the amplitude of the noise used to corrupt the data. This trend holds true regardless of whether the parametrisation of the state space is dense or sparse.
Conversely, the predictions for the solid fraction exhibit a significant sensitivity to noise levels in the sparse scenario where $M=2$. In fact, the neural model is trained using data from only two observations of solid fractions, $\phi^{(0)}$ and $\phi^{(T)}$. Thus, the synthetic noise added does not have a distribution with zero mean and its influence is exacerbated.
This result suggests that, when dealing with relatively large acquisition errors ($\delta$), the very small data set used in this study may not provide sufficient robustness, especially in terms of predicting the evolution of state variables.

\begin{figure}[h]
\centering
\begin{subfigure}{0.258\textwidth}
\centering
\includegraphics[width=0.9\textwidth]{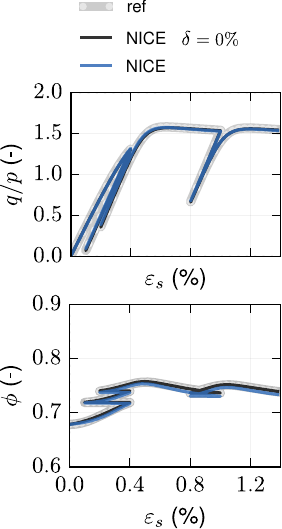}
\caption{\centering $\delta=1\%$}
\label{fig:noise_REMa}
\end{subfigure}
\hspace{1pt}
\begin{subfigure}{0.23\textwidth}
\centering
\includegraphics[width=0.9\textwidth]{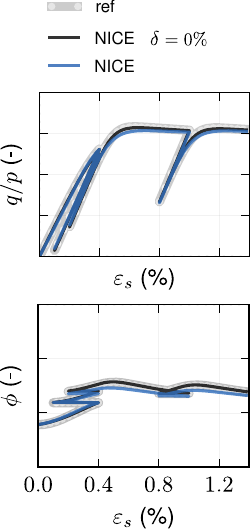}
\caption{\centering $\delta=2.5\%$}
\label{fig:noise_REMb}
\end{subfigure}
\hspace{1pt}
\begin{subfigure}{0.23\textwidth}
\centering
\includegraphics[width=0.9\textwidth]{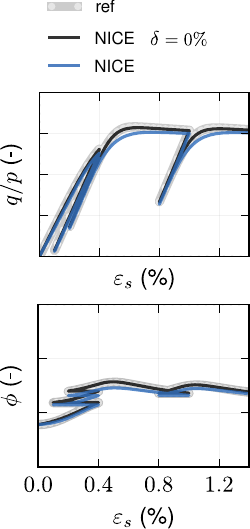}
\caption{\centering $\delta=5\%$}
\label{fig:noise_REMc}
\end{subfigure}
\hspace{1pt}
\begin{subfigure}{0.231\textwidth}
\centering
\includegraphics[width=0.9\textwidth]{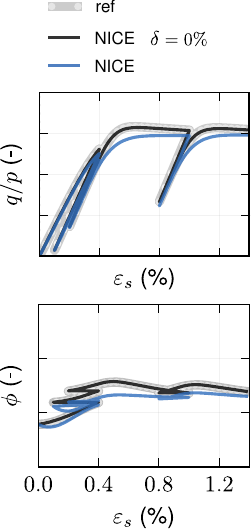}
\caption{\centering $\delta=10\%$}
\label{fig:noise_REMd}
\end{subfigure}
\caption{Influence of corrupted training data for a dense sampling of the solid fraction ($M=N$): predictions at inference in terms of the stress ratio ($q/p$, top) and solid fraction ($\phi$, bottom) for an undrained triaxial with random loading protocol. The material is prepared with initial solid fraction $\phi^{\mathrm{init}}=0.5$ and isotropically compressed up to a confining pressure of $p^{(0)}=5500$ kPa. The amplitude of the synthetic noise ($\delta$) added to the values of the stress and the state variables composing the training and validation sets varies from left to right.}
\label{fig:noise_REM}
\end{figure}

\begin{figure}[h]
\centering
\begin{subfigure}{0.258\textwidth}
\centering
\includegraphics[width=0.9\textwidth]{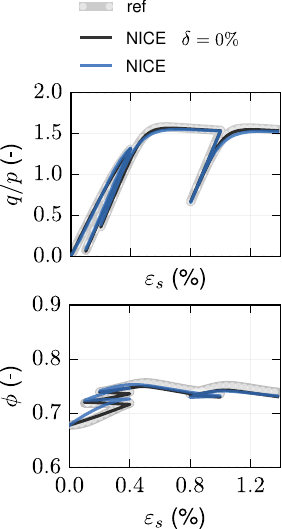}
\caption{\centering $\delta=1\%$}
\label{fig:noise_RE2a}
\end{subfigure}
\hspace{1pt}
\begin{subfigure}{0.23\textwidth}
\centering
\includegraphics[width=0.9\textwidth]{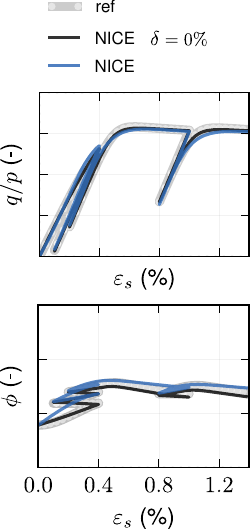}
\caption{\centering $\delta=2.5\%$}
\label{fig:noise_RE2b}
\end{subfigure}
\hspace{1pt}
\begin{subfigure}{0.23\textwidth}
\centering
\includegraphics[width=0.9\textwidth]{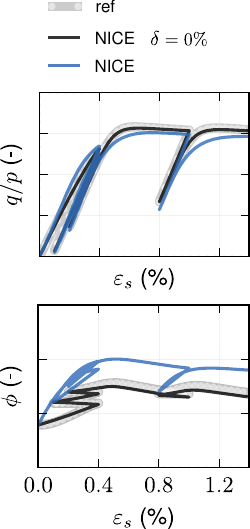}
\caption{\centering $\delta=5\%$}
\label{fig:noise_RE2c}
\end{subfigure}
\hspace{1pt}
\begin{subfigure}{0.23\textwidth}
\centering
\includegraphics[width=0.9\textwidth]{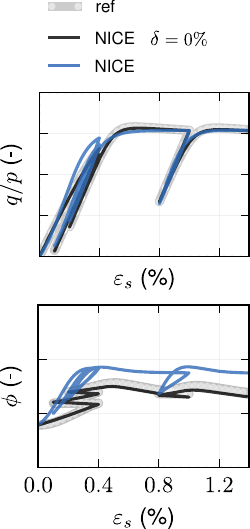}
\caption{\centering $\delta=10\%$}
\label{fig:noise_RE2d}
\end{subfigure}
\caption{Influence of corrupted training data for a sparse sampling of the solid fraction ($M=2$): predictions at inference in terms of the stress ratio ($q/p$, top) and solid fraction ($\phi$, bottom) for an undrained triaxial with random loading protocol. The material is prepared with initial solid fraction $\phi^{\mathrm{init}}=0.5$ and isotropically compressed up to a confining pressure of $p^{(0)}=5500$ kPa. The amplitude of the synthetic noise ($\delta$) added to the values of the stress and the state variables composing the training and validation sets varies from left to right.}
\label{fig:noise_RE2}
\end{figure}

Figures \ref{fig:noise_REM} and \ref{fig:noise_RE2} present the predictions for an unobserved cyclic loading path for the dense and sparse acquisition configurations, respectively. In the first case ($M=N$), the results demonstrate a high degree of accuracy in the neural model predictions. We observe a slight underestimation in the predicted values of stress and solid fraction only when subjected to high noise levels, when $\delta=10\%$.

In the case of $M=2$, the method still manages to provide stress predictions that qualitatively match the ground truth, even when noise is introduced. However, the predictions for the solid fraction, as noted earlier, are inevitably affected by the noise, leading to a less accurate representation of this variable under noisy conditions.

\section{CONCLUSIONS}
\label{sec:conclusions}
\noindent 
Despite the recent advancements in data-driven approaches, one previously unresolved issue lies in the fact that these methods requires large amounts of data. Previous neural network methods in the field of constitutive modelling therefore fail to deal with real \emph{small data} from physical experimental measurements, which can only retrieve limited partial and incomplete observations of the material state.

In this contribution, we developed a novel deep learning approach to address the issue of small data by building upon the solution of the initial value problem describing the time evolution of the material state. Inspired by recent works on neural differential equations \citep{chen2018neural}, the proposed approach blends artificial neural networks with time integration and learn time-continuous evolution equations. The approach also embraces the philosophy provided by the thermodynamics-based neural networks \citep{masi2021thermodynamics} and enables the identification of constitutive models that are always consistent with thermodynamics.

Numerical benchmarks showcased the benefits and accuracy of the proposed methodology. First, it removes the need of ``measuring'' state variables such as the elastic (or, equivalently, the plastic) strain, which cannot truly be measured or retrieved from real experiments, at least without taking gross assumptions. Second, the approach has demonstrated ability to learn accurate and robust constitutive models even in presence of generally complex model materials, and given only sparse sampling and corrupted measurement data.

Within this work, the nature of the state variables for a given material was assumed to be known. In more general cases, the same method could be advanced further to identify either: (\emph{i}) a predominant set of variables based on an initial, larger set of data, by enforcing parsimony (\emph{cf.} \citep{flaschel2023automated}), or (\emph{ii}) surrogate (hidden) state variables from the sparse acquisitions of the microscopic material state in conjunction with dimensionality reduction techniques (\emph{cf.} \citep{masistefanou2022,vlassis2022geometric}).

This work advocates the possibility of building data-driven models that require substantially smaller data sets for learning constitutive equations compared to the majority of incremental approaches. However, further exploration of the capabilities and limitations of the proposed framework are the objective of future work.
Nonetheless, the same approach can be generalised for other classes of materials beyond those considered here. In particular, it would be worthwhile to dedicate future studies on assessing whether materials with a much larger set of state variables can still be learned using only partial and limited information. This can be interesting especially for those state variables that are difficult to be collected from experiments, for instance, the granular temperature \citep{mariano2020multi,maranic2021granular} and the force-fabric in discrete media \citep{kawamoto2018all}. Similarly, future studies should be dedicated on testing the capabilities of the formulation in handling complex phenomena such as rate-dependent (\emph{e.g.} \cite{alaei2021hydrodynamic,riley2023constitutive}) and non-isothermal processes.

The newly developed formulation of neural integration for constitutive equations represents an important step to overcome the issues associated with data scarcity. Applications of the presented methodology in presence of real, experimental observations should be the ultimate objective of future works. There, accounting for additional requirements --stemming from physical (\emph{e.g.} hydrodynamics, \citep{landau2013statistical,einav2018hydrodynamic}) and mathematical \cite{germain1973method,halphen1975materiaux} considerations, including Lebesgue integration to accommodate discontinuities -- could enable further reduction of the amount of small data required for the machine \emph{learning} of even more complex material behaviours.

\section*{CODE AND DATA AVAILABILITY}
\noindent Codes and data accompanying this manuscript are publicly available in the GitHub repository in \cite{nice} (see also \href{https://github.com/filippo-masi/NICE}{{github.com/filippo-masi/NICE}}).

\section*{DECLARATION OF COMPETING INTEREST}
\noindent The authors declare that they have no known competing financial interests or personal relationships that could have appeared to influence the work reported in this paper.

\section*{ACKNOWLEDGEMENTS}
\noindent The authors would like to acknowledge the support of the Australia Research Council (ARC) under the Discovery Projects scheme (Grant agreement ID DP220101164: ``Physics-informed hydrodynamic model for clay across scales'').

\appendix

\section{APPENDIX. COMPARISON WITH THE INCREMENTAL DATA-FITTING FORMULATION}
\label{appendixA}
\noindent For the sake of completeness, the newly proposed neural integration for constitutive modelling formulation is compared with the common incremental formulation, briefly presented in subsection \ref{par:incremental}, to which we refer as \emph{neural incremental formulation}.

In doing so, we leverage the same thermodynamic framework presented in subsection \ref{par:thermodynamics} and train the finite difference approximation of the evolution equations, network $\bm{h}_{\bm{\theta}}$, and internal energy network, $\widehat{u}_{\bm{\omega}}$. However, in contrast with the neural integral constitutive equations, the training of the neural operator $\bm{h}_{\bm{\theta}}$ is not performed through the minimisation of the error associated with the solution of the initial value problem (\ref{eq:IVP_general_S}), but rather through the minimisation of the error between the predicted and ground-truth finite difference approximation of the rate of change of the state variables (as proposed in \citep{masi2023evolution}).

Note that the neural incremental formulation is conceptually simpler with respect to the one proposed in this work and is thus characterised by reduced computational complexity \citep{melchers2023comparison}. Yet, in the presence of small data, the computation of the rates or increments of the variables is inevitably subjected to high errors associated with the approximation of the time derivatives $\dot{\mathcal{X}}$ from the knowledge of sparse and time-discrete values of the state, $\mathcal{X}^{(m)}$.

The learning process for the neural incremental formulation consists of the minimisation of the following loss function, in analogy with Eq. (\ref{eq:loss})
\begin{equation}
\mathcal{L} = \mathcal{L}_{\bm{\sigma}} +\mathcal{L}_{\dot{\mathbf{z}}} + \mathcal{L}_{d}+ \lVert \mathbf{r}^{(0)}_{\bm{\varepsilon}^{e}} \rVert_2^2 + \lambda \Big(\lVert \bm{\theta} \rVert_2 +\lVert \bm{\omega} \rVert_2\Big),
\end{equation}
where
\begin{equation}
\mathcal{L}_{\dot{\mathbf{z}}}=\dfrac{1}{M-1} \sum_{m = 1}^{M-1} \Big( \frac{1}{\Delta t}\bm{h}_{\bm{\theta}}\left(\mathcal{X}^{(m)},\dot{\bm{\varepsilon}}^{(m)} \right) - \dot{\mathbf{z}}^{(m)} \Big)^2, \quad \Delta t = t^m-t^{m-1}
\end{equation}
with $\mathbf{r}^{(0)}_{\bm{\varepsilon}^{e}}$, $\mathcal{L}_{\bm{\sigma}}$, and $\mathcal{L}_d$ prescribed by Eqs. (\ref{eq:initial_ee0}, \ref{eq:loss_stress}) and (\ref{eq:loss_dissipation}). In order to compute the evolution equations of state variables, the incremental formulation requires the knowledge of the rates of those state variables $\{t^m,\dot{\mathbf{z}}^{(m)}\}_{m=0}^{M-1}$. The latter are computed relying on a first-order Euler scheme using the original $M$ (temporal) acquisitions of the state variables, in contrast with the newly proposed formulation. Notice that one could opt for a more accurate scheme, however when data are scarce it might not be always possible to do so.

After training, the time evolution of the material response under prescribed loading paths is obtained through the solution of the corresponding initial value problem (as proposed in \citep{masi2023evolution}) defined by $\bm{h}_{\bm{\theta}}$ and initial conditions $\mathcal{X}(0)=\mathcal{X}^{(0)}$.

To keep the same amount of information assumed for the training of the formulation involving the neural integration for constitutive equations, we do not prescribe here any \emph{a priori} values for the elastic strain. In contrast with the proposed approach, where only the initial value $\bm{\varepsilon}^{e(0)}$ must be identified through the solution of Eq. (\ref{eq:initial_ee0}), the neural incremental formulation demands the knowledge of the values of the elastic strain at all time points at which both the stress and the evolution equations are computed, without prescribing \emph{a priori} statements. This is achieved by generalising Eq. (\ref{eq:initial_ee0}) for all times $t^n$ and finding the corresponding $\bm{\varepsilon}^{e(n)}$ using gradient descents.

\subsection{Elasto-plastic media}
\label{appendixA1}
\noindent First consider the case of a material model with a single state variable ($\bm{\varepsilon}^e$) -- for which the results using the new formulation are described in subsection \ref{subsec:DPu} -- with those obtained using the neural incremental data-fitting formulation. Accordingly, we keep the same network typologies, parameters, initialisation strategy, and data sets. The neural incremental data-fitting approach is thus used to learn from noise-free measurement data and corrupted ones (\emph{cf.} subsection \ref{subsec:DPu}).

\begin{figure}[h]
\centering
\begin{subfigure}{0.3\textwidth}
\centering
\includegraphics[height=0.3\textheight]{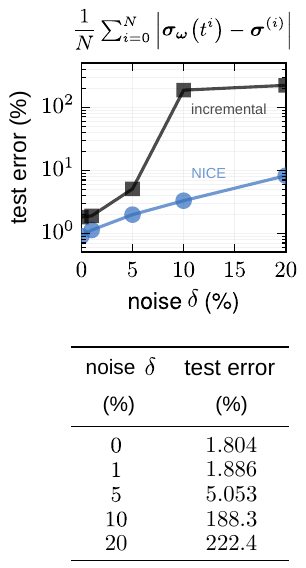}
\caption{\centering test: mean absolute error}
\label{fig:appendixA_1a}
\end{subfigure}
\begin{subfigure}{0.66\textwidth}
\centering
\includegraphics[height=0.3\textheight]{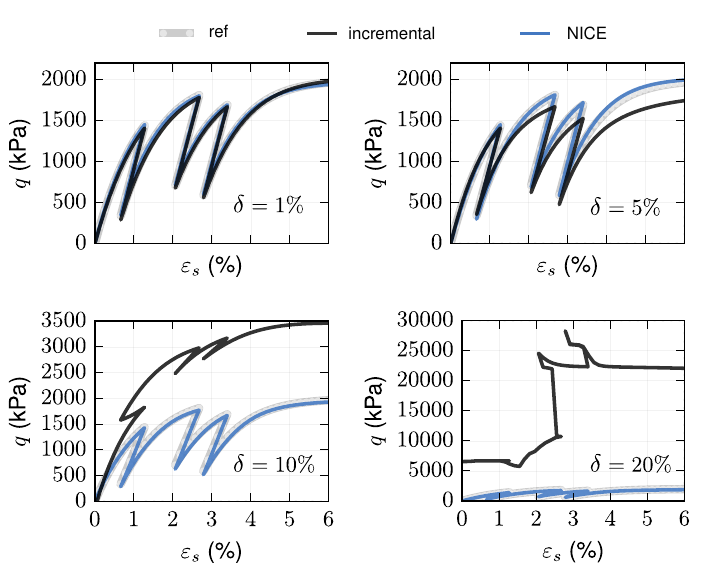}
\caption{\centering inference: stress predictions}
\label{fig:appendixA_1b}
\end{subfigure}
\caption{Conventionally incremental formulation versus the proposed approach (NICE). (a) Mean absolute percentage error of the network predictions with respect to the un-noisy test set and (b) predictions for a drained triaxial cyclical loading at constant pressure, at varying of the synthetic noise ($\delta$) added to the training and validation sets.}
\label{fig:appendixA_1}
\end{figure}

The results in terms of the error with respect to the test set and predictions at inference are presented in Figure \ref{fig:appendixA_1}. In particular, Figure \ref{fig:appendixA_1a} presents the mean absolute percentage error of the stress predictions for the test set for different levels of synthetic noise ($\delta$, \emph{cf.} subsection \ref{subsec:DPu}) in the training and validation sets. It is worth noticing that the neural incremental and integral approaches yield comparable accuracies in the case of noise-free data, with a mean absolute percentage error equal to 1.8\% for the former and to 0.9\% for the latter (\emph{cf.} Figure \ref{fig:pred_noiseDP}). However, as the noise amplitude increases, the incremental formulation is not able to learn the underlying constitutive model and its predictions are characterised by errors as high as $\approx 10\div 200\%$. This is particularly evident by comparing the results at inference with the neural integration for constitutive equations for the same cyclic loading protocols presented in Figure \ref{fig:appendixA_1b}. For $\delta \geq 5\%$, the classical approach simply fails.

\subsection{Elasto-plastic porous media}
\label{appendixA2}

\noindent Following the same framework, next consider the case of an elasto-plastic, porous material, with a total number of three state variables. We leverage, once more, the same data sets, neural network architecture, and training hyper-parameters as those detailed in subsection \ref{subsec:MCBu}.

For the sake of completeness, consider exactly the same two scenarios: (\emph{i}) a dense sampling of the solid fraction, $M=N$, and (\emph{ii}) a very sparse one, with $M=2$. The first case thus correspond to a rather more difficult problem to the one presented in \ref{appendixA1}, but where the only difficulty is given by the absence of observations for the elastic strain. The second one additionally considers data scarcity.

\begin{figure}[h]
\centering
\begin{subfigure}{0.485\textwidth}
\centering
\includegraphics[width=0.9\linewidth]{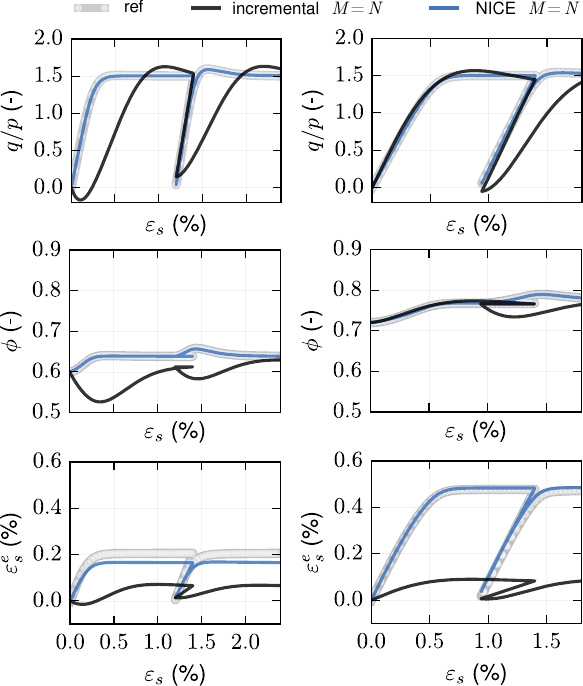}
\caption{\centering $\phi^{\mathrm{init}}=0.5$, $p^{(0)}=2500, 8500$ kPa (left, right)}
\label{fig:RE_rate_MNa}
\end{subfigure}
\hspace{1pt}
\begin{subfigure}{0.495\textwidth}
\centering
\includegraphics[width=0.9\linewidth]{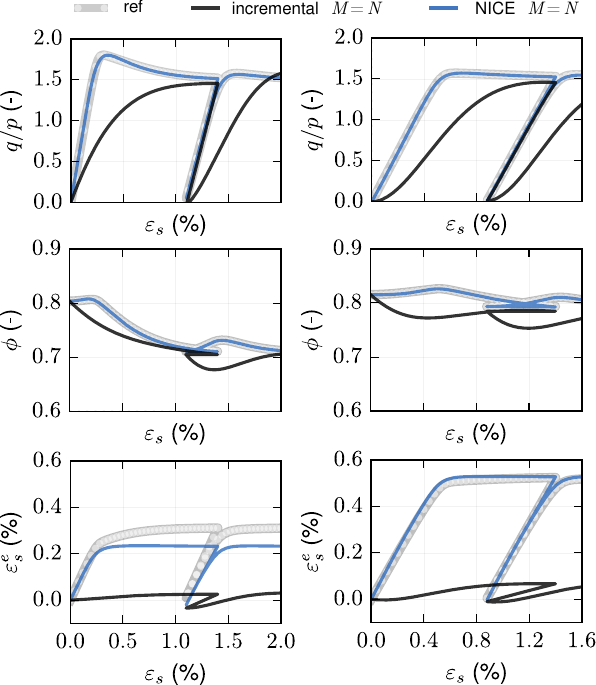}
\caption{\centering $\phi^{\mathrm{init}}=0.8$, $p^{(0)}=2500, 8500$ kPa (left, right)}
\label{fig:RE_rate_MNb}
\end{subfigure}
\caption{Conventionally incremental formulation with dense acquisition of the state variables ($M=N$) versus the proposed approach (NICE): predictions in terms of the stress ($p/q$, top), solid fraction ($\phi$, middle), and deviatoric elastic strain ($\varepsilon_s^e$, bottom) for undrained triaxial compression loading-unloading-reloading cycles with unobserved loading protocol. Material prepared with initial solid density (a) $\phi^{\mathrm{init}}=0.5$ and (b) $\phi^{\mathrm{init}}=0.8$, respectively, and isotropically compressed up to different confining pressures, $p^{(0)}=2500, 8500$ kPa.}
\label{fig:RE_rate_MN}
\end{figure}

\begin{figure}[h]
\centering
\begin{subfigure}{0.49\textwidth}
\centering
\includegraphics[width=0.9\linewidth]{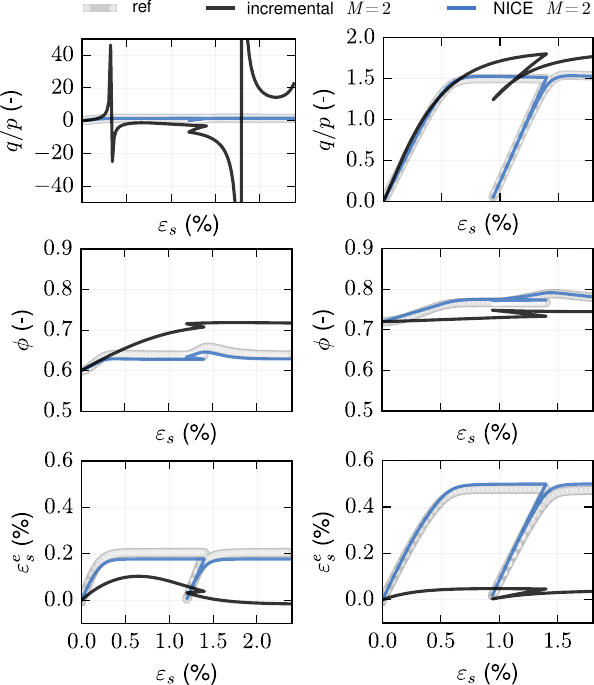}
\caption{\centering $\phi^{\mathrm{init}}=0.5$, $p^{(0)}=2500, 8500$ kPa (left, right)}
\label{fig:RE_rate_M2a}
\end{subfigure}
\hspace{1pt}
\begin{subfigure}{0.492\textwidth}
\centering
\includegraphics[width=0.9\linewidth]{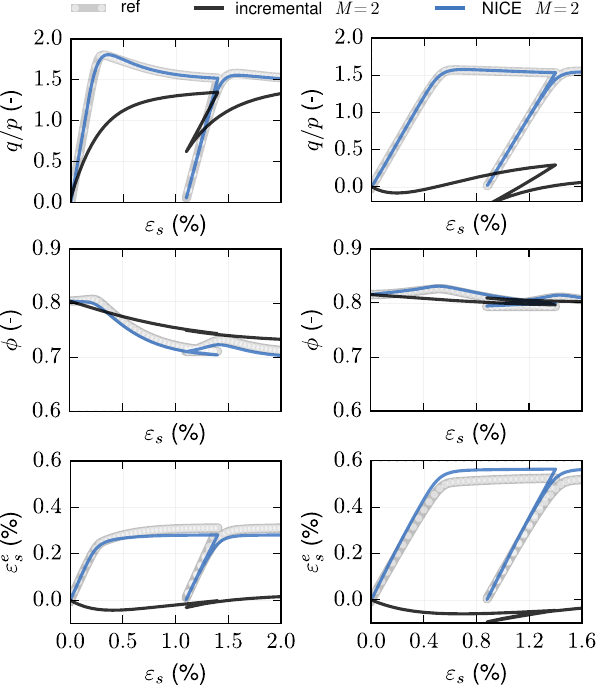}
\caption{\centering $\phi^{\mathrm{init}}=0.8$, $p^{(0)}=2500, 8500$ kPa (left, right)}
\label{fig:RE_rate_M2b}
\end{subfigure}
\caption{Conventionally incremental formulation with sparse acquisition of the state variables ($M=2$) versus the proposed approach (NICE): predictions in terms of the stress ($p/q$, top), solid fraction ($\phi$, middle), and deviatoric elastic strain ($\varepsilon_s^e$, bottom) for an undrained triaxial compression loading-unloading-reloading cycles with unobserved loading protocol. Material prepared with initial solid density (a) $\phi^{\mathrm{init}}=0.5$ and (b) $\phi^{\mathrm{init}}=0.8$, respectively, and isotropically compressed up to different confining pressures, $p^{(0)}=2500, 8500$ kPa.}
\label{fig:RE_rate_M2}
\end{figure}

The predictions of the neural incremental formulation are presented in Figures \ref{fig:RE_rate_MN} and \ref{fig:RE_rate_M2} for an unobserved loading path, and are compared with those provided by the newly proposed approach, as well as with the ground-truth solution. The classical incremental data-fitting formulation clearly fails regardless of the sampling frequency for the solid fraction. The difference with the newly proposed approach only lies in the formulation (incremental versus integral). These results demonstrate that the thermodynamic structure, presented in subsection \ref{par:thermodynamics} and kept in the neural incremental approach, is not a sufficient to learn from small data.

The reason lies in the fact that the neural incremental formulation cannot correctly learn the evolution of both the elastic strain and the solid fraction. Whilst thermodynamics enables the network to distinguish the nature of the two variables, the lack of observations for the elastic strain reveals the fallacy of the traditional incremental formulation.

Given the unrealistic predictions obtained for the incremental formulation, it seems reasonable to infer that in the case of corrupted measurement data, the predictions can only become worst.

\end{document}